\documentclass[journal]{IEEEtran}
\usepackage{amsmath,amsfonts}
\usepackage{algorithmic}
\usepackage{algorithm}
\usepackage{array}
\usepackage[caption=false,font=normalsize,labelfont=sf,textfont=sf]{subfig}
\usepackage{textcomp}
\usepackage{stfloats}
\usepackage{url}
\usepackage{verbatim}
\usepackage{graphicx}
\usepackage{cite}
\usepackage{amssymb}
\usepackage{multirow}
\usepackage{makecell}
\usepackage{color, soul}
\usepackage{booktabs}
\usepackage{pifont}
\usepackage{hyperref}
\hypersetup{
	colorlinks=true,
	linkcolor=blue,
	urlcolor=red,
	citecolor=blue,
}
\usepackage[table]{xcolor} 
\usepackage{colortbl} 
\usepackage{marvosym}
\newcommand{\highlight}[2]{\colorbox{#1!20}{\textbf{#2}}}

\begin{document}

\title{SHARE: A Fully Unsupervised Framework for Single Hyperspectral Image Restoration}

\author{Jiangwei Xie$^*$, \and  Zhang Wen$^*$, \and Mike Davies, 
\and Dongdong Chen\textsuperscript{\Letter}
\thanks{
$*$: Equal Contribution, \Letter: corresponding author (\text{d.chen@hw.ac.uk}). 
Jiangwei Xie, Zhang Wen, Dongdong Chen are with the School of Mathematical and Computer Sciences, Heriot-Watt University, Edinburgh, UK. Mike Davies is with the School of Engineering, University of Edinburgh, UK.}
}


\maketitle

\begin{abstract}
Hyperspectral image (HSI) restoration is a fundamental challenge in computational imaging and computer vision. It involves ill-posed inverse problems, such as inpainting and super-resolution. Although deep learning methods have transformed the field through data-driven learning, their effectiveness hinges on access to meticulously curated ground-truth datasets. This fundamentally restricts their applicability in real-world scenarios where such data is unavailable.  This paper presents SHARE (Single Hyperspectral Image Restoration with Equivariance), a fully unsupervised framework that unifies geometric equivariance principles with low-rank spectral modelling to eliminate the need for ground truth. SHARE's core concept is to exploit the intrinsic invariance of hyperspectral structures under differentiable geometric transformations (e.g. rotations and scaling) to derive self-supervision signals through equivariance consistency constraints. Our novel Dynamic Adaptive Spectral Attention (DASA) module further enhances this paradigm shift by explicitly encoding the global low-rank property of HSI and adaptively refining local spectral-spatial correlations through learnable attention mechanisms. Extensive experiments on HSI inpainting and super-resolution tasks demonstrate the effectiveness of SHARE. Our method outperforms many state-of-the-art unsupervised approaches and achieves performance comparable to that of supervised methods. We hope that our approach will shed new light on HSI restoration and broader scientific imaging scenarios. The code will be released at \url{https://github.com/xuwayyy/SHARE}.

\end{abstract}


\section{Introduction}
\IEEEPARstart{H}{yperspectral} image (HSI) restoration aims to recover a clear HSI from a potentially noisy and degraded HSI that follows a known linear degradation model. Since real-world hyperspectral imaging always suffers from missing pixels or low resolution \cite{degrade-source}, it plays a crucial role in practice, enabling a wide range of applications in various fields, such as agriculture \cite{agriculture-1}, mineral exploration \cite{mineral-2}, and environmental monitoring \cite{envir-2}. 

From a variational perspective, solving HSI restorations involves minimising the negative log-likelihood:
\begin{align}
    &\min_x \{-\log p(x|y)\} \tag{1a} \label{eq2a} \\
    \propto &\min_x \{-\log p(y|x) - \log p(x)\} \tag{1b} \label{eq2b}\\
    \propto &\min_x \{\mathcal{L}(\mathcal{H}(x),y) + \mathcal{R}(x)\} \tag{1c} \label{eq2c} 
\end{align}
where $\mathcal{L}_{\mathcal{H}(x)}$ is the data fidelity term and $\mathcal{R}(x)$ is the regularization term such as smoothness, symmetry and sparsity.

Conventional HSI restoration methods primarily focus on exploiting abundant spectral information, such as sparsity-based methods \cite{low-rank-2015} and total variation (TV)-based methods \cite{ tv-2017}. However, these methods often suffer from high computational complexity and difficulty in recovering fine-grained information, especially in challenging scenarios. In contrast to the construction of hand-crafted priors, learning-based methods \cite{DeNet-learning-2018-1} \cite{DualTransformer-learning-2023-5} \cite{ISDiff-learning-2024-6} \cite{diffusion-learning-2024-7} \cite{latent-diff-learning-2024-8}\cite{zhao2024image} directly build a mapping from observation \( \{y\} \) to ground truth \( \{x\} \) through a data-driven strategy. Among them, supervised methods mainly use a CNN, a Vision Transformer (VIT) \cite{VIT} or a Diffusion \cite{DDPM} network as a backbone and learn from paired $\{(x, y)\}$ data sets. Although these methods have achieved promising results in popular restoration tasks such as denoising, super-resolution, and inpainting, the construction of large HSI datasets remains impractical due to the high cost of imaging systems.

Moreover, since HSI datasets are rare, these approaches often suffer from poor domain generalisation. Recently, self-supervised methods \cite{DHIP} \cite{PnP-DIP} \cite{self-super-two-stage-2021} \cite{self-super-assisted-2024} \cite{hir-diff-self-2024} have gained popularity due to their ability to work without paired data sets. However, most of these methods still rely on auxiliary supervised learning to provide pre-trained restoration models, indicating the use of ground truth during training. When there is a significant domain shift between the source and target data, their performance still degrades dramatically. Although methods such as \cite{DIP}, \cite{DHIP} and \cite{PnP-DIP} can operate in a fully unsupervised manner, they often suffer from a significant performance gap compared to supervised approaches. Furthermore, the robustness of these methods is further challenged by the fact that images captured by real-world sensors are often noisy and incomplete.

To address these issues, we propose a novel fully unsupervised HSI restoration framework: \textbf{S}ingle \textbf{H}yperspectral Im\textbf{A}ge \textbf{R}estoration with \textbf{E}quivariance (\textbf{SHARE}). SHARE is based on the fact that image signals remain invariant under a group of transformations. Unlike previous self-supervised approaches, SHARE imposes mild invariance priors, allowing learning from single image at test time. Moreover, our approach is robust to noisy observations by incorporating the Stein Unbiased Risk Estimator (SURE) \cite{SURE-loss}, which adheres to real-world hyperspectral imaging conditions. Furthermore, we design a U-shaped architectural network in $f_\theta$ to further exploit informative features.

Overall, our contribution can be summarized as follows:
\begin{itemize}
    \item We propose a novel fully unsupervised HSI restoration framework \textbf{SHARE}, which exploits the invariance property of images and learning at test time on the single image.
    
    \item We incorporate the SURE to SHARE, enhancing its robustness under real-world sensor imaging conditions and enabling a robust self-supervised HSI restoration method.
    
    \item Extensive experiments on HSI inpainting and super-resolution show that SHARE achieves state-of-the-art performance among unsupervised methods and comparable performance to supervised methods.
\end{itemize}

The rest of this paper is organized as follows. Section~\ref{sec:related-work} reviews the recent works related to HSI restoration and equivariant imaging. Section~\ref{sec:method} presents our proposed method for HSI restoration. Section~\ref{sec:experiment} presents the extensive experimental and ablation studies. We conclude our paper in section~\ref{sec:conclusion}.

\section{Related Work}
\label{sec:related-work}

\subsection{HSI Restoration}

Prior-based methods always use hand-crafted priors and iteratively reconstruct a clean HSI image. They include sparsity  \cite{sparsty-sr-2014-eccv} \cite{sparsty-sr-2019-tip}, low-rank approximation \cite{low-rank-2018-jstar} are most commonly used. TV regularization \cite{tv-2017} is also always used for compressed sensing \cite{tv-compress-2020-tip} and denoising \cite{tv-2018-denoising-tgrs}. Although these methods always exploit both spatial and spectral information, they always suffer from difficult optimization and poor performance in complex scenarios.

Learning-based methods \cite{DeNet-learning-2018-1}  \cite{DualTransformer-learning-2023-5} \cite{ISDiff-learning-2024-6} \cite{diffusion-learning-2024-7}, on the other hand, commonly tend to adopt a CNN or Vision Transformer backbone. Chang et al. \cite{DeNet-learning-2018-1} proposed to use CNN to restore HSI. Wei et al. \cite{QRNN3D-learning-2021-tnnls} proposed a 3D quasi-recurrent network (QRNN3D) for HSI denoising and inpainting. Cai et al. \cite{coase-to-fine-learning-2022-4} build a coarse-to-fine transformer architecture to restore HSI. Li et al. \cite{latent-diff-learning-2024-8} incorporates Diffusion and VIT to further improve performance. However, these methods need to be trained on large datasets, which are hard to collect in the real world. Although data augmentation alleviates this problem, these methods always struggle with poor performance in high-resolution and noisy conditions. 

Recently, self-supervised learning has gained significant attention. However, in the HSI domain, self-supervised methods still show a significant performance gap compared to their supervised counterparts. Several studies \cite{self-super-two-stage-2021} \cite{self-super-TSBSR} \cite{self-super-assisted-2024} \cite{hir-diff-self-2024} have attempted to bridge this gap by pre-training a restoration model in a supervised manner to gain generalization ability, followed by fine-tuning the model in an unsupervised setting. While these methods show promising results, they still rely on ground truth data to support the learning process. By contrast, methods such as \cite{DHIP} and \cite{PnP-DIP}, which are the closest to our proposed method in principle, use DIP \cite{DIP} to iteratively reconstruct clean images from observations alone. However, they often rely on heuristic architectural design and lack well-defined design principles.
\begin{figure}
    \centering
    \includegraphics[width=1\linewidth]{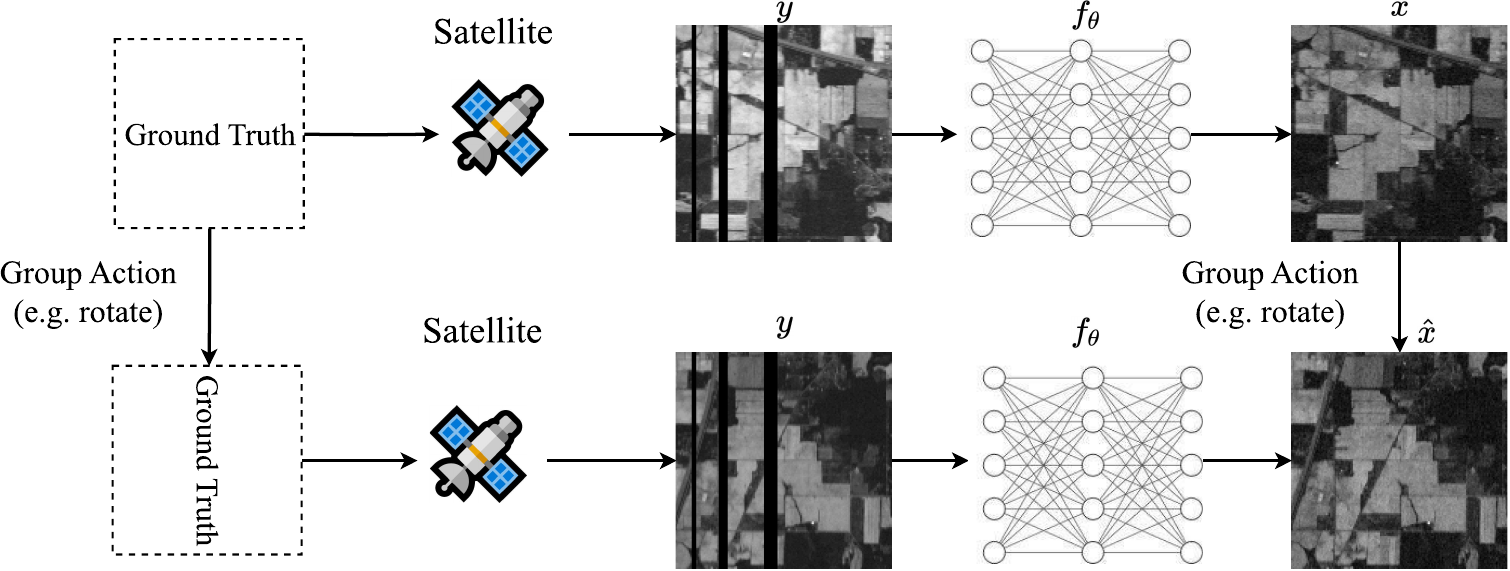}
    \caption{An illustration of learning with equivariance in the HSI inpainting task. The reconstructed image after applying a transformation should be equivalent to the restoration of the initially transformed and masked image.}
    \label{fig:ei-in-inpainting}
\end{figure}

\subsection{Equivariant Imaging}
Equivariant Imaging (EI) \cite{EI} \cite{REI} \cite{SPM} is the state-of-the-art fully unsupervised framework for solving inverse imaging problems and has been successfully applied to a variety of computational imaging and low-level vision tasks such as Cryo-EM image reconstruction \cite{liu2024overcoming}, image fusion \cite{EMMA}, super-resolution \cite{ei-sr}, and pansharpening \cite{ei-pansharpen}. EI uses the group invariance properties inherent in natural signals to learn a reconstruction function using only partial measurement data. It assumes that the original signal set has symmetry properties and remains invariant under a group of transformations - such as rotation, translation and scaling. By embedding such invariance priors, EI ensures that the entire image processing pipeline (from acquisition to reconstruction) exhibits transformation equivariance. 

As demonstrated in \cite{EI}\cite{JMLR}, it is possible to learn to image from measurements alone using image signals' intrinsic symmetry  and low dimensional structures. To answer these key factors, we carefully select \textit{Shift} and \textit{Scale} group actions and explicitly incorporate a DASA module with low dimension, thereby enforcing the HSI recovery from only measurement HSI.  

We note that the recent works also allows learning with equivariance \cite{Hyper-EI} \cite{ei-pansharpen}. The difference between them and the proposed one twofold: (1) the related work either lack robustness to noise \cite{Hyper-EI} or focus on a single task \cite{ei-pansharpen}, whereas this paper aims to provide a more generalized framweork that can simultaneously address both inpainting and super-resoluton. (2) we proposed a DASA module which leverages the low-dimensional (low-rank) property of HSI signals, which significantly improves EI for HSI image restoration.

\section{Method}
\label{sec:method}

\subsection{Problem Statement}

We formulate the single-image HSI restoration problem for inpainting and super-resolution tasks. Given \textbf{only one} noisy and incomplete observation $y\in\mathbb{R}^{c\times h \times w}$ generated by a known linear degradation model $\mathcal{H}$ acting on the ground truth $x\in\mathbb{R}^{C\times H \times W}$ with additive noise $\epsilon$, where $C$ is the bands number, $H$ and $W$ represent the spatial size, the challenge is to recover $x$ from only $y$. Since the training set contains only one image, SHARE performs several iterations and directly outputs the clean HSI in a \emph{zero-shot manner}, which is adhere to the fact that HSI dataset always contain only one image.

\begin{itemize}
    \item \textbf{HSI Inpainting:}
In real-world satellite imaging, missing stripes may occur in certain areas due to the platform’s inability to consistently maintain its programmed attitude, limited access to accurate digital elevation maps (DEMs), or misalignment between the flight path and the hyperspectral sensors \cite{Hyper-EI}.
For the inpainting task, $\mathcal{H}$ is a binary diagonal matrix and $y$ is obtained through:
\begin{equation}
    y = \mathcal{H} \odot x + \epsilon
\label{eq:inpainting-degrade}
\end{equation}
where $\odot$ represents the Hadamard product. 

\item \textbf{HSI Super-resolution:} Due to limited solar irradiance, there is an inherent trade-off between spatial and spectral resolution in HSI; prioritizing spectral resolution often results in reduced spatial resolution \cite{HSI-SR-motivation}. For the super-resolution task, $\mathcal{H}$ is modeled by a Gaussian blur kernel $\Phi$ and downsampling factor $s$:
\begin{equation}
    y = (x* \Phi)\downarrow_{s} + \epsilon
\label{eq:sr-degrade}
\end{equation}
where $*$ is convolution operator, $\downarrow_s$ means downsampling.  
\end{itemize}

Clearly, this single observation constraint is more in line with real HSI restoration scenarios where the acquisition is always only one noisy and corrupted image, but it renders supervised learning approaches inapplicable. In the next section we show how our SHARE solves these problems.

\begin{figure*}
    \centering
    \includegraphics[width=0.7\linewidth]{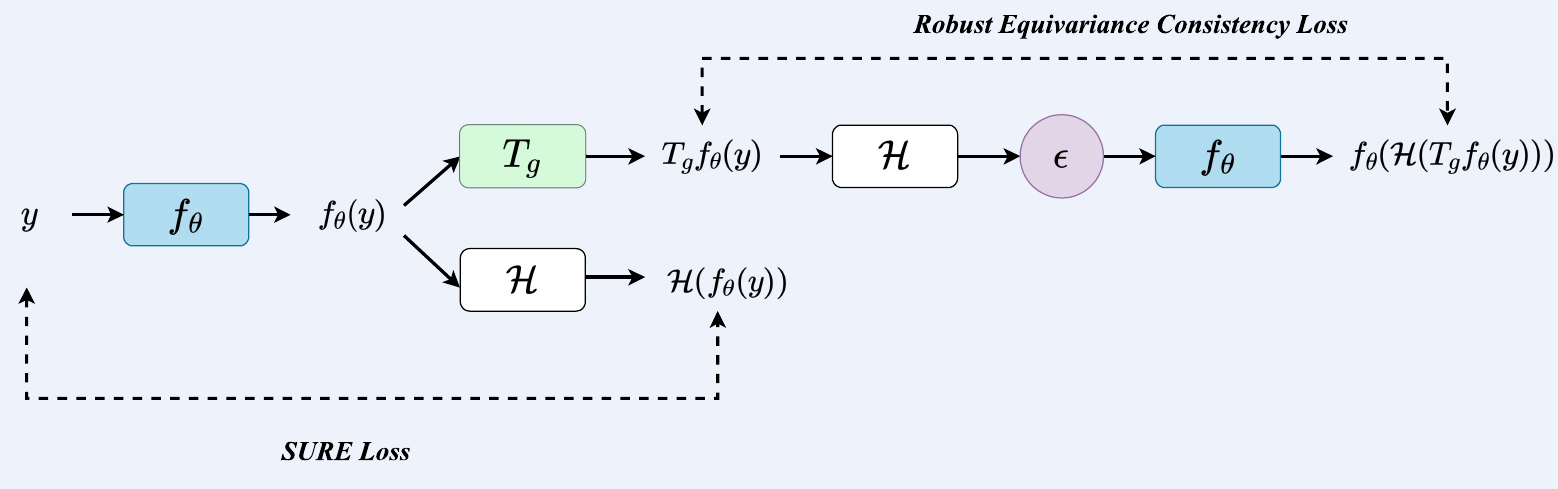}
    \caption{The overall pipeline of SHARE. Given only one noisy observation $y$, it is first passed through the inverse mapping $f_\theta$ to obtain a restored estimate $f_\theta(y)$. This estimate is then forwarded through the physics model $\mathcal{H}$ to compute $\mathcal{H}(f_\theta(y))$, which is compared with the observation $y$ to calculate the SURE loss. Simultaneously, $f_\theta(y)$ undergoes a transformation $T_g$, followed by the physics model, noise corruption, and inverse mapping, to compute the robust equivariance consistency loss.}
    \label{fig:overview}
\end{figure*}

\subsection{Single HSI Restoration with Equivariance}
In supervised learning, the goal is to minimize the error between the predicted clean image and the corresponding ground truth. This typically involves optimizing a loss function, such as the mean squared error between the reconstructed image \( f(y) \) and the ground truth \( x \). However, obtaining a large amount of ground truth data is often challenging in real-world hyperspectral imaging. Therefore, in unsupervised learning (ground truth-free) scenarios, the dataset consists only of degraded images, making the availability of ground truth \( x \) non-existent. Consequently, the training loss must be redefined to eliminate the dependence on ground truth data.

\noindent\textbf{Learning Without Robustness}

\subsubsection{Measurement Consistency}

Let $f_\theta: y\rightarrow x$ be the reconstruction function. A straightforward approach in unsupervised learning is to enforce measurement consistency (MC) by minimizing the discrepancy between observation \( y \) and the degradation of the reconstructed image \( \mathcal{H}f_{\theta}(y) \). This can be expressed as:
\begin{equation}
    \mathcal{L}_{mc}=\mathbb{E}_{y}\left[\|\mathcal{H}(f_{\theta }(y))-y\|_{2}^2\right]
\label{eq:mc}
\end{equation}

However, the estimation of $f_{\theta}$ cannot be achieved by minimizing the error between $\mathcal{H}(f_{\theta}(y))$ and $y$ for two reasons: First, we don't have the null space information of $\mathcal{H}$ w.r.t. the ground truth, i.e., the clean components in the ground truth image, but only have access to the information in its range space, i.e., the degraded components in $y$, and we need to learn more information beyond these range space components \cite{EI}\cite{DDN}. Furthermore, such a naive estimation is not robust. In real-world sensor imaging conditions, the captured images are always noisy, which further limits the restoration. To address these challenges, we follow \cite{EI} and utilize an equivariance module to learn clean components so that the ground truth $\{x\}$ can be recovered from its raw observation $\{y\}$, and propose to use SURE instead of vanilla MC.

\begin{figure*}
    \centering
    \includegraphics[width=1\linewidth]{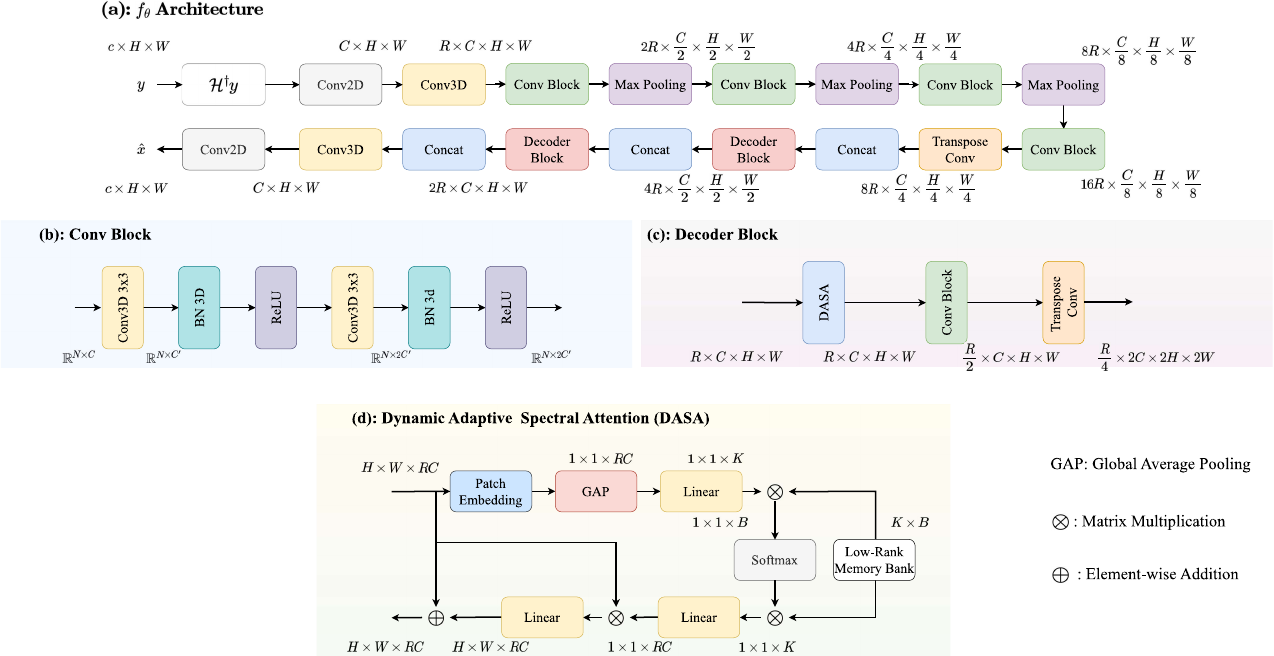}
    \caption{(a) The overall architecture of the inverse mapping $f_\theta$, which adopts a U-shaped design. (b) The structure of the convolutional block. It doubles the input's channel dimension in the encoding phase and halves it in the decoding phase. (c) The architecture of the decoder block consists of a DASA module, a convolutional block, and a transposed convolution layer. (d) The internal structure of the proposed DASA module.}
    \label{fig:module}
\end{figure*}

\subsubsection{Equivariance Consistency}
We start with the observation that image signals often remain invariant under a group of transformations. This concept can be intuitively illustrated by the example of satellite imaging, where the camera continuously moves and captures the same target area from different \emph{viewpoints}. Despite these variations in viewpoint, the semantic content of the image remains unchanged. Such scenarios naturally lead to transformation invariance, including rotation, translation, scaling, etc.

As pioneered in \cite{EI}, system equivariance could enable the unknown signal set to be recovered from its compressed measurements by interacting with the group action $T_g$ and the physics $\mathcal{H}$. In particular, the restoration function $f_\theta$ can be learned from only measuremnets if the intersection of submanifolds $\cap_g \mathcal{H}_g^{-1}y$ contains a single signal $x$, where $\mathcal{H}_g=\mathcal{H}\circ T_g$. In principle, if the image set $\{x\}$ is invariant with respect to a transformation group $\{T_g\}$, then the whole degrade-restore system $f_\theta\circ \mathcal{H}$ should be equivariant with respect to these transformations, i.e.
\begin{equation}
    f_\theta \bigl(\mathcal{H}(T_{g} x)\bigr) = T_{g} \bigl( f_\theta(\mathcal{H}(x)) \bigr).
\label{eq:equi}
\end{equation}

Since the ground truth ${x}$ is not available, inspired by \cite{EI}, we adopt the following ground-truth-free equivariance constraint:
\begin{equation}
    f_\theta(\mathcal{H}\bigl(T_{g} f_{\theta}(y)\bigr)) = T_{g}f_{\theta}(y).
\label{eq:ei}
\end{equation}
where the right-hand side follows from the fact $\mathcal{H}(x) = y$ under zero noise condition. Note that on the left-hand side, the first $f_\theta$ takes as input the transformed prediction passed through the measurement operator $\mathcal{H}$, whereas the second $f_\theta$ operates directly on the original observation $y$. Despite this difference in input, the same network $f_\theta$ is used on both sides. According to \cite{JMLR}, Eq.~\ref{eq:ei} does not require $\mathcal{H}$ or $f_\theta$ to be invariant under $T_g$, but the composition $f_\theta \circ \mathcal{H}$ to be equivariant under $T_g$. Fig.~\ref{fig:ei-in-inpainting} demonstrates the idea of Eq.~\ref{eq:equi} under the HSI inpainting task. The reconstructed image after rotation should be equal to the rotated and then reconstructed image.

\noindent\textbf{Learning With Robustness}

In most real-world situations, observed measurements are corrupted by noise. For example, measurements typically contain Gaussian noise, i.e. $y \sim \mathcal{N}(\mathcal{H}(x), I\sigma^2)$. To derive a robust self-supervised HSI restoration method, following \cite{REI}, we modify Eq.~\ref{eq:mc} and Eq.~\ref{eq:ei}, respectively. The modified noise robust framework is illustrated in Fig.~\ref{fig:overview}.

\subsubsection{Robust Equivariance Constraint}
To ensure consistency with the distribution in the testing phase (i.e., noisy images), the degraded images used during training must also contain noise, i.e. robust equivariance consistency \cite{REI}:
\begin{equation}
        f_\theta(\tilde{y}) = T_{g}f_{\theta}(y)
\label{eq:rei}
\end{equation}
where $\tilde{y}\sim \mathcal{N}(\mathcal{H}\bigl(T_{g} f_{\theta}(y)\bigr), I\sigma^2)$ is the new noisy measurement derived from the transformed reconstruction $T_{g}f_{\theta}(y)$.  Note in the absence of noise, Eq.~\ref{eq:rei} is equivalent to Eq.~\ref{eq:ei}.

\subsubsection{SURE Loss}
Recently, Chen et al. introduced a noise-robust variant of the original EI by incorporating Stein's Unbiased Risk Estimate (SURE) function \cite{SURE-loss} to estimate the clean measurement consistency\cite{REI}. In this paper, we are interested in investigating the potential of SURE in solving inverse problems in the hyperspectral domain. In general, the SURE loss provides an unbiased estimator of the supervised mean square loss between the noisy observation $y$ and its corresponding ground truth $x$ \cite{SURE-loss}. Under the assumption of Gaussian noise, the SURE loss is formulated as follows.
\begin{equation}
    \mathcal{L}_{sure} = ||y - \mathcal{H} f_\theta(y)||^2_2 - \sigma^2 + 2\sigma^2\nabla f_\theta(y)
\label{eq:sure-1}
\end{equation}
where $\sigma$ is the standard deviation of the noise. The first term in Eq.~\ref{eq:sure-1} penalizes the discrepancy between the noisy observations and the restored results, reflecting the bias of the reconstruction neural network, while the last term penalizes the variance, promoting stability in the denoising process.

Since direct computation of the divergence term is often intractable, \cite{monteCarlo-SURE} proposed a Monte Carlo method to estimate the divergence:
\begin{align}
    \nabla f_\theta(y) &= \lim_{\tau\rightarrow0} \mathbb{E}_{b} \{\frac{b^\top}{\tau} (\mathcal{H}(f_\theta(y+\tau b)) - \mathcal{H}(f_\theta(y)))  \} \\
    &\approx \frac{b^\top}{\tau} (\mathcal{H}(f_\theta(y+\tau b)) - \mathcal{H}(f_\theta(y)))
    \label{Eq: monte-carlo-sure}
\end{align}
where $b$ is an i.i.d. random vector with zero mean and unit variance. Thus the divergence can be efficiently approximated by sampling $b \sim \mathcal{N}(0, I)$ and using a small scalar $\tau$. 

Finally, if the noise distribution is Gaussian, the SURE loss has the following expression, where the divergence term is estimated via Monte Carlo approximation.
\begin{align}
    \mathcal{L}_{sure} = &~ \|y - \mathcal{H}f_\theta(y)\|_2^2 - \sigma^2 \notag\\
    &+ \frac{2\sigma^2}{\tau} b^\top\big(\mathcal{H}(f_\theta(y+\tau b)) - \mathcal{H}(f_\theta(y))\big)
\label{eq:final-gaussian-sure}
\end{align}

Note that in the ablation study we also evaluate our SHARE model under Poisson and mixed Gaussian-Poisson noise settings. A detailed derivation of the SURE loss function for all noise conditions can be found in \cite{REI}. In the new paragraph we will introduce the architecture of $f_\theta$.

\subsection{Inverse Mapping $f_\theta$}

\subsubsection{Overview of $f_\theta$}
The architecture of $f_\theta$ is illustrated in Fig.~\ref{fig:module}(a). The main network adopts a U-shaped structure. Given a noisy observation $y \in \mathbb{R}^{c \times H \times W}$, it is first passed through a linear inverse module $\mathcal{H}^\dagger y$ and a 2D convolution to increase the channel dimension from $c$ to $C$. The resulting feature map is denoted as $z_0 \in \mathbb{R}^{C \times H \times W}$.

Inspired by previous success of 3D convolution in a various of HSI tasks \cite{3DConv-motivation-1} \cite{QRNN3D-learning-2021-tnnls} \cite{3DConv-motivation-2}, we also consider 3D convolution in $f_\theta$. After acquiring $z_0$, it is lifted to a 4D tensor $z_1 \in \mathbb{R}^{R \times C \times H \times W}$ by a 3D convolution, where $R$ represents the spectral depth. This tensor enters the encoder path. In the $i$-th encoder stage, the feature $e_i$ is obtained by a convolution block that doubles the depth dimension: $e_i \in \mathbb{R}^{2^i R \times C/2^i \times H/2^i \times W/2^i}$. Each convolution block consists of 3D convolution, batch normalisation and ReLU activation. After each block, 3D max pooling is applied to reduce spatial and spectral resolution.

In the decoder path, a transpose convolution is first applied to the feature map to double the spatial and spectral dimensions and halve the depth dimension. The output is then concatenated with the corresponding encoder feature $e_i$ along the depth axis. This combined tensor is passed through a decoder block that includes a spectral attention module, a convolution block, and another transpose convolution. Each decoder block reduces the depth dimension by a factor of 4 while doubling the spatial and spectral resolution.

Finally, the decoder output is passed through a 3D convolution to reduce the depth to 1, followed by a 2D convolution that restores the channel dimension to the original $c$, resulting in the final restored HSI image $\hat{x} \in \mathbb{R}^{c \times H \times W}$.

\subsubsection{Dynamic Adaptive Spectral Attention}

Early studies \cite{HSI-Unmixing-endmember} have suggested that the spectral signatures in HSI can often be modeled as a linear combination of a limited number of endmembers, reflecting a low-dimensional subspace structure in the spectral domain \cite{chen2017dmc}. This insight underpins a variety of classical unmixing and reconstruction approaches. Building on this foundation, recent methods have leveraged the low-rank nature of HSI data to enhance restoration tasks, which can be adopted to alleviate the computational complexity of 3D convolutions while maintaining strong channel-wise feature extraction capabilities. To this end, inspired by \cite{li-spectral-low-rank-unit}, we propose a dynamic adaptive spectral attention (DASA) module. 

Our method uses a low-rank memory bank with a user-defined rank, which plays a similar role to the reduction ratio in \cite{SENet} and exploits the inherent low-rank structure of HSI. The memory bank can be interpreted as a learnable dictionary of global low-rank spectral vectors and dynamically adapts to different local features.

The structure of our proposed DASA module is shown in Fig.~\ref{fig:module} (d). Given a 4D input tensor, we first remove the depth dimension and transform it into a 3D tensor. To capture spectral-spatial dependencies, the features are partitioned into non-overlapping cube patches of size $P \times P \times C$. Following \cite{SENet}, a global average pooling operation is applied within each patch to produce a compact spectral descriptor $X_c \in \mathbb{R}^{1 \times 1 \times RC}$. Then, a linear projection layer maps $X_c$ to a low-rank representation $X_k \in \mathbb{R}^{1 \times 1 \times K}$.

\begin{figure*}
    \centering
    \includegraphics[width=1\linewidth]{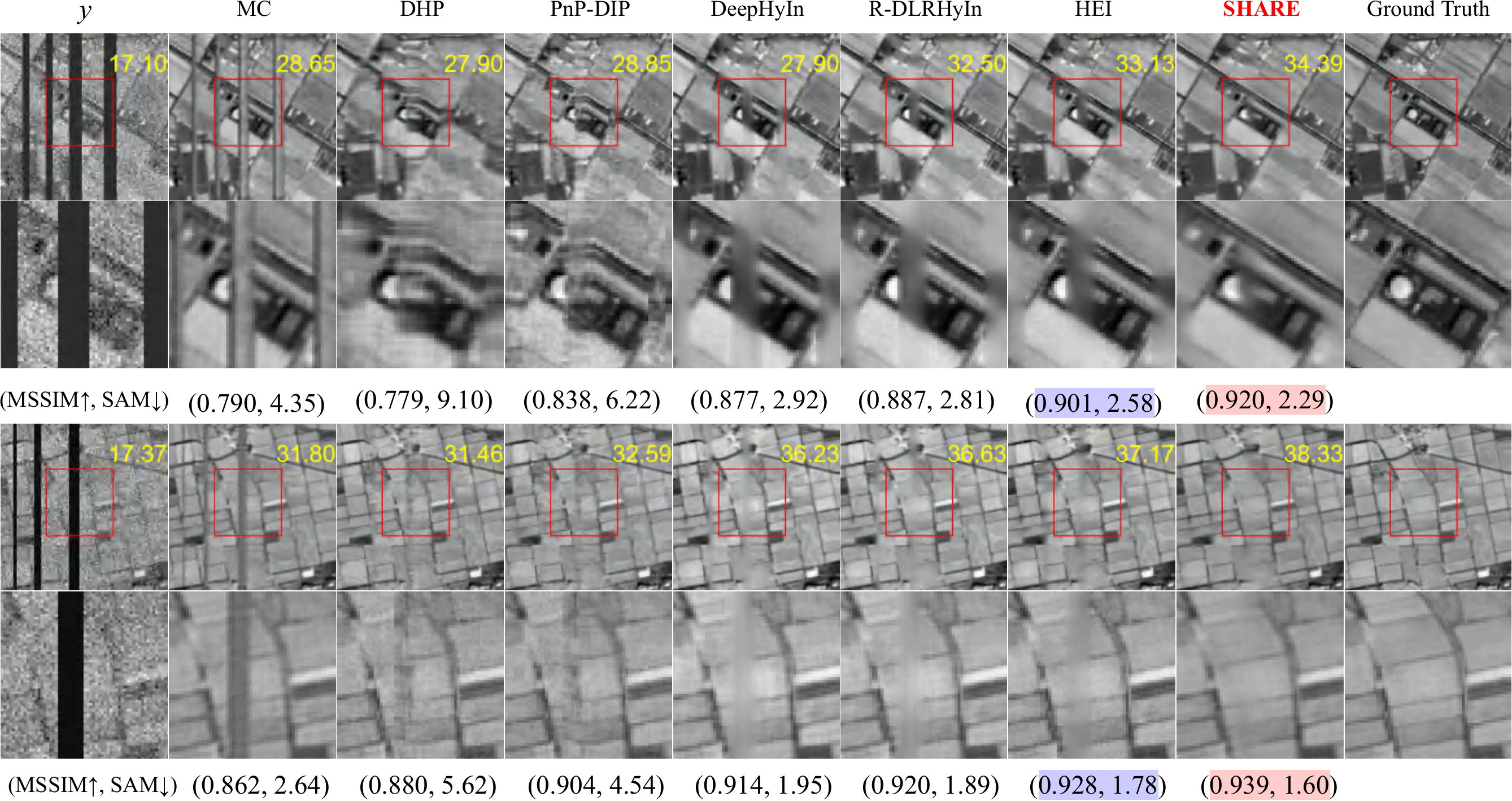}
    \caption{Inpainting results of different methods on \textbf{Chikusei} dataset. MPSNR values are shown on the top left, MSSIM and SAM values are shown below each image. The best and second-best values are marked in \highlight{red}{red} and \highlight{blue}{blue}, respectively. Band 90 for visualization. Please zoom in for a better view.}
    \label{fig:inpainting-chikusei}
\end{figure*}

Next, the memory bank computes the similarity between $X_k$ and the stored low-rank matrix $M \in \mathbb{R}^{K \times B}$ where $B$ is the number of low-rank matrix in memory bank, resulting in a weight vector $I \in \mathbb{R}^{1 \times 1 \times B}$. In implementation, $M$ is a learnable parameter and is updated via backpropagation. These weights are used to aggregate vectors from the memory bank and obtain an enhanced low-rank representation $X_l \in \mathbb{R}^{1 \times 1 \times K}$, as formulated below:
\begin{align} 
        I &= \text{Softmax}(X_k M)  \\ 
        X_l &= I M^\top. 
\end{align}

Notably, instead of projecting the entire image into a global rank-$K$ subspace, the proposed memory-based mechanism allows each local patch to attend to global low-rank bases. This allows effective global-local interaction and adaptively improves the spectral representation. The resulting vector $X_l$ preserves the most informative features of the original HSI and dynamically adapts to different inputs through content-aware weighting.

Finally, $X_l$ is projected back to the original channel dimension and reshaped to match the spatial dimensions of the input. A residual connection is then applied to produce the final output of the DASA module.

\subsection{Loss Function}
To guarantee the robust equivariance consistency in Eq.~\ref{eq:rei}, we use $\mathcal{L}_{rec}$ to minimize their error:
\begin{equation}\label{eq:share_rec}
    \mathcal{L}_{rec} = \mathbb{E}_{y, g} [||T_gf_\theta(y) - f_\theta(\tilde{y_i})||_2^2]
\end{equation}

The final loss function for SHARE is the combination of the SURE loss and the robust equivariance consistency loss. Its ground truth free expression is as follows
\begin{equation}
    \mathcal{L}_{SHARE} = \mathcal{L}_{sure} + 
\alpha \mathcal{L}_{rec}
\label{eq:share}
\end{equation}
where $\alpha$ is a trade-off parameter, with these two terms, SHARE is robust to noise and learns fully without ground truth.

\begin{figure*}
    \centering
    \includegraphics[width=1\linewidth]{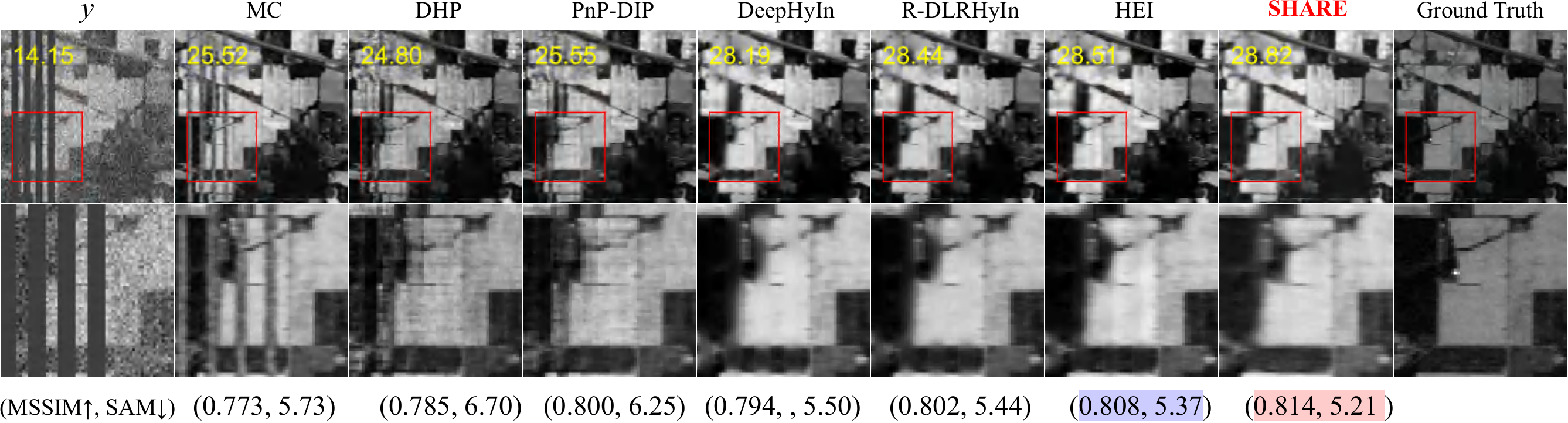}
    \caption{Inpainting results of different methods on \textbf{Indian-Pines} dataset. MPSNR values are shown on the top left, MSSIM and SAM values are shown below each image. The best and second-best values are marked in \highlight{red}{red} and \highlight{blue}{blue}, respectively. Band 149 for visualization. Please zoom in for a better view.}
    \label{fig:inpainting-indian}
\end{figure*}

\section{Experiments}
\label{sec:experiment}

\subsection{Implementation Details}
We implement all experiments based on Pytorch and the DeepInverse library \cite{deepinv}, using a single NVIDIA GeForce RTX 3090 GPU. For training details. We use the Adam optimiser with an initial learning rate of $1e^{-3}$ and a cosine scheduler to $1e^{-4}$ within 2000 epochs. The $\alpha$ in Eq.~\ref{eq:share} is set to 1 and the $\tau$ in Eq.~\ref{eq:final-gaussian-sure} is set to 0.01. The rank in the memory bank is $K=4$ and $B=256$. As suggested by \cite{EI} and \cite{ei-sr}, we employ Shift and Scale as transformation groups for inpainting and super-resolution, respectively, given their demonstrated effectiveness for the associated measurement operator $\mathcal{H}$. For both inpainting and super-resolution, the observations are corrupted by Gaussian noise with zero mean and $25 \ / \ 255$ standard deviation.

\begin{table*}[htbp]
    \centering
        \caption{Comparison of different methods on \textbf{Chikusei} and \textbf{Indian-Pines} datasets. We report the average values of each metric on four masks and all images for each dataset. The \highlight{red}{red} and \highlight{blue}{Blue} markers indicate the best and second-best values among all methods. / means can not compute metric value.}
    \begin{tabular}{c|ccc ccc}
        \toprule
         \multirow{2}{*}{Method} & \multicolumn{3}{c}{Chikusei Dataset} & \multicolumn{3}{c}{Indian Pines Dataset} \\
         \cmidrule{2-7}
         & MPSNR $\uparrow$ & MSSIM $\uparrow$ & SAM $\downarrow$ & MPSNR $\uparrow$ & MSSIM $\uparrow$ & SAM $\downarrow$ \\
        \hline
        $\mathcal{H}^\dagger y$ & 16.96 & 0.163 & / & 13.485 & 0.279 & /  \\
        MC  & 29.59 & 0.795 & 4.68 & 24.52 & 0.716  & 6.42 
        \\
        DHP \cite{DHIP} & 28.93 & 0.828 & 6.98 & 24.39 & 0.756 & 7.36\\ 
        PnP-DIP \cite{PnP-DIP}& 29.97 &0.865 &5.57 & 24.94 & 0.777 & 6.82 \\
        DeepHyIn \cite{DeepHyIn}& 33.52 & 0.888 &3.56 &27.64 & 0.771 & 5.76 \\
        R-DLRHyIn \cite{R-DLRHyIn}& 33.97 & 0.898 & 3.07 & 27.82 & 0.775 & 5.70 \\
        HEI \cite{Hyper-EI} & \highlight{blue}{34.18} & \highlight{blue}{0.901} & \highlight{blue}{2.93} &  \highlight{blue}{28.03} & \highlight{blue}{0.789} & \highlight{blue}{5.57}\\
        \textbf{SHARE} & \highlight{red}{35.12} & \highlight{red}{0.917} & \highlight{red}{2.56} & \highlight{red}{28.54} & \highlight{red}{0.801} & \highlight{red}{5.41}\\
         \bottomrule
    \end{tabular}
    \label{tab:inpainting}
\end{table*}

\begin{table*}
    \caption{Comparison of different methods on \textbf{Cave: fake and real beers ms} dataset. The \highlight{red}{red} and \highlight{blue}{Blue} markers indicate the best and second-best values among all methods.}
    \centering
    \begin{tabular}{c|ccc c ccc c  ccc}
        \toprule
        {Method} & \multicolumn{3}{c}{Downsample Rate $\times 2$} & & \multicolumn{3}{c}{Downsample Rate $\times 4$}& & \multicolumn{3}{c}{Downsample Rate $\times8$}  \\
        \cmidrule{2-12}
        
        & MPSNR $\uparrow$ & MSSIM $\uparrow$ & SAM $\downarrow$ &  & MPSNR $\uparrow$ & MSSIM $\uparrow$ & SAM $\downarrow$  & & MPSNR$\uparrow$ & MSSIM$\uparrow$ & SAM 
        $\downarrow$  \\
        
        \cline{1-12}
        Bicubic & 21.38 & 0.180 & 18.45 & & 21.07 & 0.265 &  18.22 &  & 20.65 & 0.450 & 18.05   \\

        DHP \cite{DHIP} & 27.53 & 0.538 & 7.90 & & 26.13 & 0.592 & 9.00 && 19.88 & 0.182 & 17.01    \\
        PnP-DIP \cite{PnP-DIP} & 29.31 & 0.689 & 5.55 & & 26.69 & 0.594 & 6.53 && 24.15 & 0.517 & 7.03\\
        SSDL \cite{self-super-assisted-2024}&  30.45 & 0.762 & 4.48 & & 27.83 & 0.730 & 5.69 & & 25.63 & 0.699 & 6.94     \\

        MC &  27.36 & 0.547 & 8.54 & & 26.37 & 0.429 & 10.60 & & 21.88 & 0.227 & 16.96    \\

        Supervised \cite{MCNet} &  \highlight{red}{36.97} & \highlight{red}{0.961} & \highlight{red}{2.12} & & \highlight{red}{33.75} & \highlight{red}{0.927} &  \highlight{blue}{2.92} & & \highlight{blue}{29.14} & \highlight{blue}{0.884} & \highlight{blue}{4.64}    \\   
        \textbf{SHARE} & \highlight{blue}{35.60} & \highlight{blue}{0.938} & \highlight{blue}{2.64} & & \highlight{blue}{33.22} & \highlight{blue}{0.916} & \highlight{red}{2.86} & & \highlight{red}{30.46} & \highlight{red}{0.893}  & \highlight{red}{3.27}        \\ 
        \bottomrule
    \end{tabular}
    \label{tab:sr-cave-ms}
\end{table*}

\begin{figure*}
    \centering
    \includegraphics[width=1\linewidth]{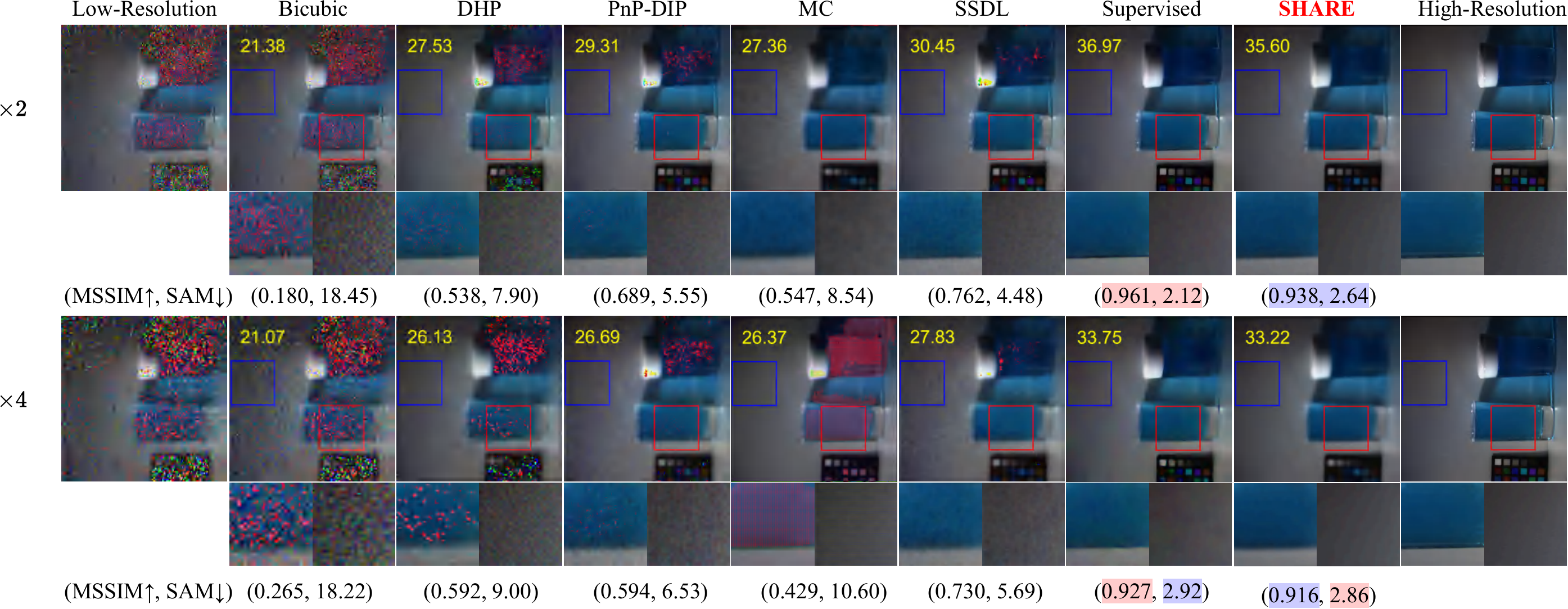}
    \caption{Super resolution results of different methods on \textbf{Cave: fake and real beers ms} dataset. MPSNR values are shown on the top left, MSSIM and SAM values are shown below each image. The best and second-best values are marked in \highlight{red}{red} and \highlight{blue}{blue}, respectively. Bands 5, 15, 25 for visualization. Please zoom in for a better view.}
    \label{fig:beers}
\end{figure*}
\subsection{Evaluation Metrics}
For both inpainting and super-resolution tasks, we select three common metrics for HSI restoration: Mean Peak Signal to Noise Ratio (MPSNR), Mean Structural Similarity Index Measure \cite{SSIM} (MSSIM), and Spectral Angle Mapper (SAM). Higher MPSNR and MSSIM values correspond to better results, while lower SAM values indicate better results.

\subsection{HSI Inpainting}
We conduct experiments on two real-world remote sensing datasets: Chikusei \cite{Chikusei} and Indian-Pines \cite{Indian-pine}. The Chikusei dataset contains 192 spectral channels. Following \cite{Hyper-EI}, we crop five $144 \times 144$ sub-images which serve as ground truth. The Indian Pines dataset was acquired by the AVIRIS sensor and contains 220 spectral bands with a spatial resolution of $145 \times 145$ pixels. We remove the last row and column of pixels to match the resolution of the Chikusei dataset and remove 20 corrupted spectral bands, preserving 200 bands. Four different mask shapes are used to corrupt the ground truth. Following \cite{Hyper-EI}, all the masks corrupt the whole bands in one column, with mask ratio $12.5\%, 23.6\%, 16.67\%, 41.67\%$. 

We compare our SHARE with conventional linear pseudo-inverse $\mathcal{H}^{\dagger} y$, six learning-based methods: DHP \cite{DHIP}, PnP-DIP \cite{PnP-DIP}, DeepHyIn \cite{DeepHyIn}, R-DLRHyIn \cite{R-DLRHyIn}, HEI \cite{Hyper-EI}, and MC. In implementation, We used the same Skip-Net \cite{SKIPNET} as the backbone for the DHP, PnP-DIP and R-DLRHyIn. These methods are all designed for single image inpainting and learn without ground truth. 

The experimental results on the Chikusei and Indian Pines HSI inpainting datasets further verify the superiority of the SHARE method. As shown in Table ~\ref{tab:inpainting}, SHARE achieves the best performance on all three metrics on the Chikusei dataset, with an MPSNR of 35.12, MSSIM of 0.917, and SAM of 2.56, clearly outperforming the previous state-of-the-art HEI (MPSNR: 34.18, MSSIM: 0.901, SAM: 2.93). On the Indian Pines dataset, SHARE also shows strong results, achieving an MPSNR of 28.54, MSSIM of 0.801 and SAM of 5.41, with particular excellence in the structurally relevant MSSIM and SAM metrics, indicating improved perceptual and structural fidelity.

\begin{table*}
    \caption{Comparison of different methods on \textbf{Cave: glass tiles ms} dataset. The \highlight{red}{red} and \highlight{blue}{Blue} markers indicate the best and second-best values among all methods.}
    \centering
    \begin{tabular}{c|ccc c ccc c  ccc}
        \toprule
        {Method} & \multicolumn{3}{c}{Downsample Rate $\times 2$} & & \multicolumn{3}{c}{Downsample Rate $\times 4$}& & \multicolumn{3}{c}{Downsample Rate $\times8$}  \\
        \cmidrule{2-12}
        
        & MPSNR $\uparrow$ & MSSIM $\uparrow$ & SAM $\downarrow$ &  & MPSNR $\uparrow$ & MSSIM $\uparrow$ & SAM $\downarrow$  & & MPSNR$\uparrow$ & MSSIM$\uparrow$ & SAM 
        $\downarrow$  \\
        
        \cline{1-12}
        Bicubic & 20.73 & 0.158 & 45.38 & & 20.10 & 0.145 & 44.75 & & 19.54 & 0.19 & 44.06  \\

        DHP \cite{DHIP}& 25.10 & 0.642 & 23.39 && 22.45 & 0.479 & 26.48 && 17.21 & 0.104 & 48.82 \\
        PnP-DIP \cite{PnP-DIP} & 25.79 & 0.760 & 20.22  && 23.04 & 0.648 & 23.38 && 20.42 & 0.319 & 32.29\\
        SSDL \cite{self-super-assisted-2024}& 28.64 & 0.776 & 12.78 & & 26.45 & 0.684 &
        14.10 & & 25.87 & 0.680 & 14.78  \\
       
        MC  &  26.81 & 0.703 & 18.51 & & 25.85 & 0.625 & 
        19.02 & & 24.17 & 0.603 & 19.25 \\
         Supervised \cite{MCNet} & \highlight{red}{32.78} & \highlight{red}{0.875} & \highlight{blue}{9.45} & & \highlight{blue}{29.15} & \highlight{red}{0.779} & \highlight{blue}{11.69} & & \highlight{blue}{26.40} & \highlight{blue}{0.667} & \highlight{blue}{15.01} \\
        \textbf{SHARE} & \highlight{blue}{31.22} & \highlight{blue}{0.854} & \highlight{red}{8.36} & & \highlight{red}{28.77} & \highlight{blue}{0.775}
        & \highlight{red}{9.17} & & \highlight{red}{26.77} & \highlight{red}{0.716} & \highlight{red}{10.49} \\ 
        \bottomrule
    \end{tabular}
    \label{tab:sr-cave-glass}
\end{table*}

It is important to note that the MC method cannot inpaint the masked regions. Instead, it only restores the known range space, which explains its relatively high MPSNR. This suggests that MC preserves pixel-level similarity in the unmasked regions, but fails to recover coherent structure in the occluded regions. In contrast, SHARE effectively restores both masked and unmasked areas, demonstrating a strong ability to model global structure and local details.

The visual results in Fig. \ref{fig:inpainting-chikusei} and Fig. \ref{fig:inpainting-indian} support these findings. SHARE produces clearer edges, more accurate textures, and perceptually pleasing results. In scenes with complex occlusions and structural damage, SHARE is able to seamlessly reconstruct missing content and maintain smooth transitions between restored and original regions, demonstrating its advantages in contextual modelling and geometric consistency. Please see Supplementary Material (SM) for more inpainting results with more mask shapes.

\subsection{HSI Super-Resolution}
We perform experiments on three popular datasets: CAVE \cite{CAVE}, Pavia University \cite{Pavia}, and Chikusei \cite{Chikusei}. For the CAVE dataset, we select 'fake and real beers ms' and 'glass tiles ms' as ground truth, with each HSI having a shape of $512\times 512\times 31$. For the Pavia University dataset, we centrally crop $320\times320\times 103$ sub-images as ground truth. For the Chikusei dataset, we centrally crop a patch of $512\times 512\times 128$. The CAVE dataset is downsampled by $\times 2$, $\times 4$ and $\times 8$, while the other two datasets are downsampled by $\times 2$ and $\times 4$. To ensure sufficient blurring and anti-aliasing, Gaussian blur is applied using kernels of sizes $15\times15$, $29\times29$, and $55\times55$ with standard deviations of $2$, $4$, and $8$ for the CAVE dataset, and half of kernel size values for the remote sensing dataset, respectively
\begin{figure*}
    \centering
    \includegraphics[width=1\linewidth]{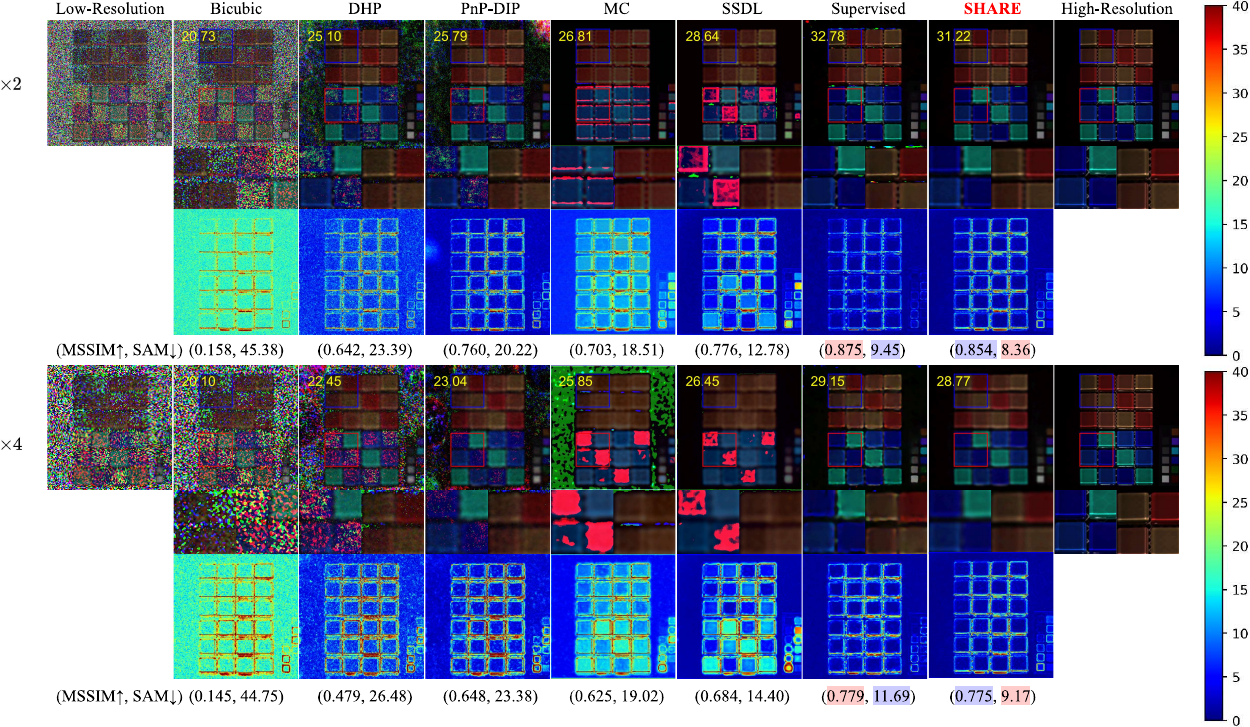}
    \caption{Super resolution results of different methods on \textbf{Cave: glass tiles ms} dataset. MPSNR values are shown on the top left, MSSIM and SAM values are shown below each image. The best and second-best values are marked in \highlight{red}{red} and \highlight{blue}{blue}, respectively. Bands 5, 15, 25 for visualization. First row: Super resolution results. Second row: Local details. Third row: Absolute error map. Please zoom in for a better view.}
    \label{fig:glass}
\end{figure*}

\begin{table*}
    \caption{Comparison of different methods on \textbf{Chikusei} dataset. The \highlight{red}{red} and \highlight{blue}{Blue} markers indicate the best and second-best values among all methods. The values are computed as the average results of two subfigures of Chikusei Dataset.}
    \centering
    \begin{tabular}{c|ccc ccc}
        \toprule
        {Method} & \multicolumn{3}{c}{Downsample Rate $\times 2$} &  \multicolumn{3}{c}{Downsample Rate $\times 4$} \\
        \cmidrule{2-7}
        & MPSNR $\uparrow$ & MSSIM $\uparrow$ & SAM $\downarrow$  & MPSNR $\uparrow$ & MSSIM $\uparrow$ & SAM $\downarrow$   \\
        \cline{1-7}
        Bicubic &  20.03 &0.110 & 23.49 & 19.85 & 0.153 & 23.61    \\
        DHP \cite{DHIP} & 28.52 & 0.692 & 6.36 & 25.22 & 0.447 & 10.16 \\ 
        PnP-DIP \cite{PnP-DIP} & 28.85 & 0.741 & 5.28 & 26.23 & 0.554 & 7.85\\ 
        SSDL \cite{self-super-assisted-2024}& \highlight{blue}{29.25} & \highlight{blue}{0.757} & \highlight{blue}{4.82} & \highlight{blue}{26.96} & \highlight{blue}{0.655} & \highlight{blue}{8.65}  \\
        MC  & 28.16 & 0.591 & 9.00 & 26.87 & 0.487 & 10.48   \\
        \textbf{SHARE} & \highlight{red}{32.64} & \highlight{red}{0.822} & \highlight{red}{4.18} & \highlight{red}{30.52} & \highlight{red}{0.733}  & \highlight{red}{5.04}  \\ 
        
        \bottomrule
    \end{tabular}
    \label{tab:sr-chikusei}
\end{table*}

We compare our SHARE method with conventional bicubic interpolation and four learning-based methods: DHP \cite{DHIP}, PnP-DIP \cite{PnP-DIP}, and SSDL \cite{self-super-assisted-2024}, and MC. We also compare SHARE with supervised methods \cite{MCNet} on the CAVE dataset. To the best of our knowledge, they are the most recent learning-based works for \emph{unsupervised single HSI super-resolution} with the exception of \cite{MCNet}. Note that \cite{MCNet} cannot be directly applied to single HSI super-resolution, so we train it on other CAVE images, using rotation, flipping and patch cropping for data augmentation. For SSDL \cite{self-super-assisted-2024}, although the authors have not open-sourced their code, their core loss function components are included in the DeepInverse library \cite{deepinv}, and the Bi-3DQRNN backbone \cite{QRNN3D-learning-2021-tnnls} is in the public domain. Therefore, we implement this method ourselves. In addition, for a fair unsupervised comparison, we remove the first stage of supervised learning in SSDL, leaving only the second stage of unsupervised learning. After this modification, all methods except the supervised one can be trained with single image.

\begin{figure*}
    \centering
    \includegraphics[width=1\linewidth]{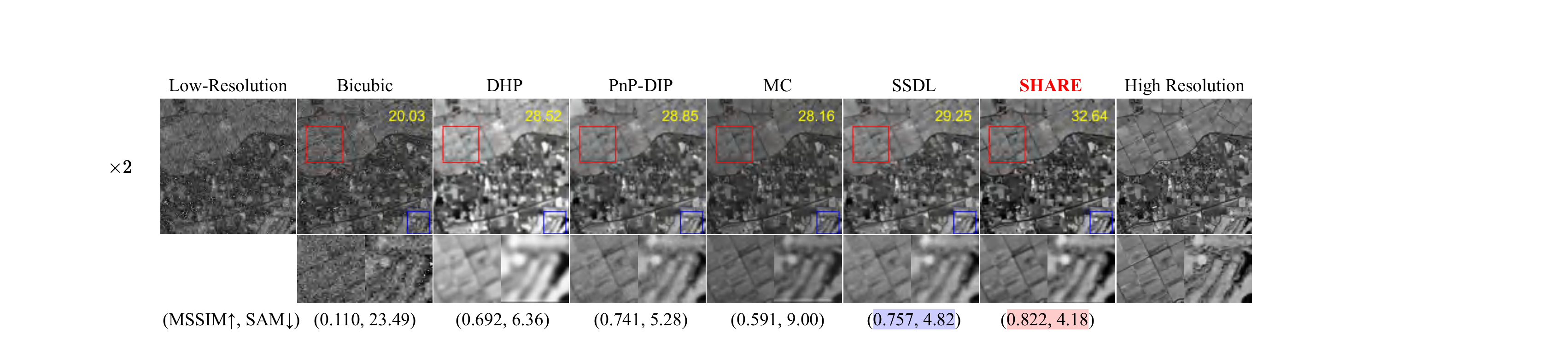}
    \caption{Super resolution results of different methods on \textbf{Chikusei} dataset. We crop two subfigures from Chikusei. MPSNR values are shown on the top left, MSSIM and SAM values are shown below each image. The best and second-best values are marked in \highlight{red}{red} and \highlight{blue}{blue}, respectively. Band 90 for visualization. Please zoom in for a better view.}
    \label{fig:chikusei-sr}
\end{figure*}

\begin{figure*}
    \centering
    \includegraphics[width=1\linewidth]{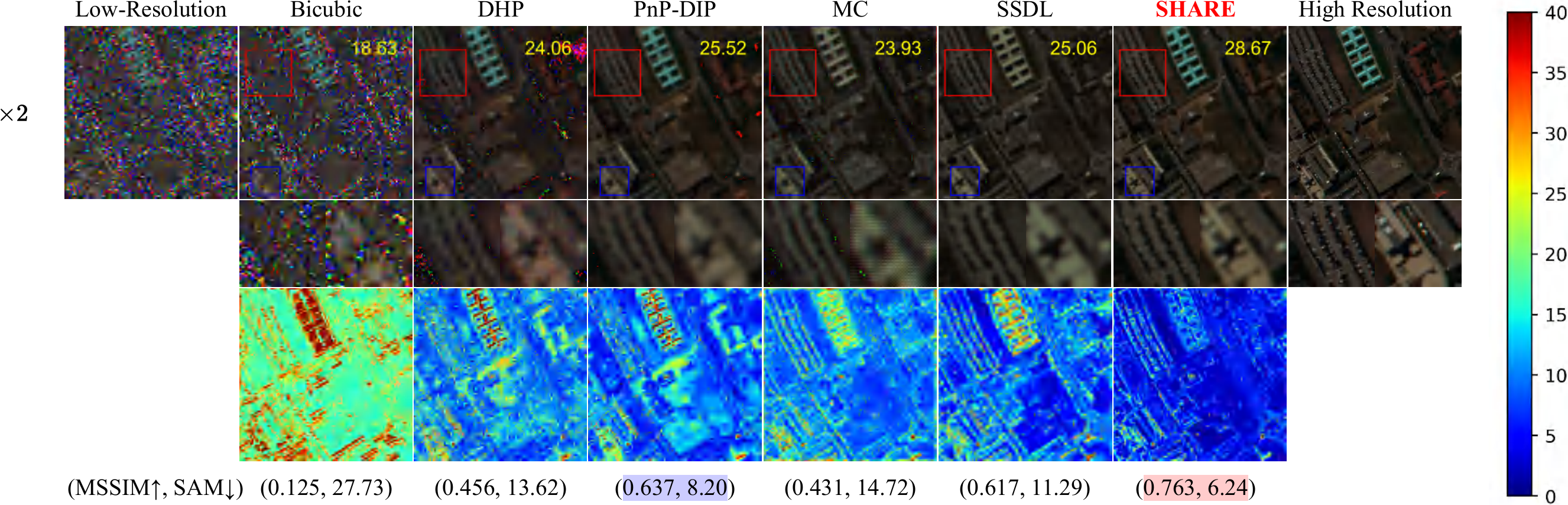}
    \caption{Super resolution results of different methods on \textbf{Pavia University} dataset. MPSNR values are shown on the top left, MSSIM and SAM values are shown below each image. The best and second-best values are marked in \highlight{red}{red} and \highlight{blue}{blue}, respectively. Bands 60, 29, 7 for visualization. First row: Super resolution results. Second row: Local details. Third row: Absolute error map. Please zoom in for a better view.}
    \label{fig:pavia}
\end{figure*}

On the CAVE dataset, we evaluate SHARE on the fake/real beers ms and glass tiles ms. For the beers dataset, the quantitative results are reported in Table~\ref{tab:sr-cave-ms}, while the corresponding visual comparisons are shown in Fig.~\ref{fig:beers}. At all dowmsampling scales, it ranks second overall but remains the strongest among unsupervised approaches, being surpassed only by the supervised method. The visual comparisons further demonstrate that SHARE preserves edge structures and fine textures—especially around the bottle boundaries and the color checker—better than traditional unsupervised methods. Despite being fully unsupervised, SHARE achieves a reconstruction quality comparable to supervised approaches.

For the glass tiles dataset, the quantitative results are provided in Table~\ref{tab:sr-cave-glass}, and the visual results are illustrated in Fig.~\ref{fig:glass}. SHARE demonstrates robust performance across all scales. At $\times4$ and $\times8$, it consistently ranks second overall but still outperforms all unsupervised baselines, reaching 26.77 MPSNR, 0.716 MSSIM and 10.49 SAM at $\times8$. We also plot the error maps to demonstrate the quality of spectral reconstruction. It is evident that SHARE outperforms other methods and achieves performance comparable to the supervised approach. Please see SM for a more detailed comparison of the high-resolution results.

The University Pavia dataset demonstrates the SHARE model, which shows significant advantages at both the $\times2$ and $\times4$ downsampling scales. As shown in Table~\ref{tab:sr-pavia}, SHARE achieves the best performance across all evaluation metrics. The visual results in Fig.~\ref{fig:pavia} further support this observation: SHARE preserves finer details, clearer edges and more accurate textures, especially in the highlighted regions (red and blue boxes), where the reconstructed images are visibly closer to the high-resolution ground truth references. Furthermore, we present the error maps of all techniques, where SHARE significantly surpasses the others in terms of visual quality. See SM for a more detailed comparison of the high-resolution results.

\begin{table*}
    \caption{Comparison of different methods on \textbf{Pavia University} dataset. The \highlight{red}{red} and \highlight{blue}{Blue} markers indicate the best and second-best values among all methods.}
    \centering
    \begin{tabular}{c|ccc ccc}
        \toprule
        {Method} & \multicolumn{3}{c}{Downsample Rate $\times 2$} &  \multicolumn{3}{c}{Downsample Rate $\times 4$} \\
        \cmidrule{2-7}
        
        & MPSNR $\uparrow$ & MSSIM $\uparrow$ & SAM $\downarrow$  & MPSNR $\uparrow$ & MSSIM $\uparrow$ & SAM $\downarrow$   \\
        
        \cline{1-7}
        Bicubic &  18.63 & 0.125 & 27.73 & 18.28 & 0.155
        & 28.13  \\

        DHP \cite{DHIP} & 24.06 & 0.456 & 13.62 & 22.62 & 0.356 & 13.56 \\
        PnP-DIP \cite{PnP-DIP} & \highlight{blue}{25.52} & \highlight{blue}{0.637} & \highlight{blue}{8.20} & 23.09 & 0.425 & \highlight{blue}{11.31}\\
        SSDL \cite{self-super-assisted-2024}& 25.06 & 0.617 & 11.29 & \highlight{blue}{23.38} & \highlight{blue}{0.508} & 12.89    \\

        MC &  23.93 & 0.431 & 14.72 & 21.87 & 0.384 & 19.19   \\
        
        \textbf{SHARE} & \highlight{red}{28.67} & \highlight{red}{0.763} & \highlight{red}{6.24} & \highlight{red}{25.77} & \highlight{red}{0.602} & \highlight{red}{7.85} \\ 
        \bottomrule
    \end{tabular}
    \label{tab:sr-pavia}
\end{table*}

Finally, the performance of our model on the Chikusei dataset also shows advantages over other methods. As shown in Table~\ref{tab:sr-chikusei}, SHARE achieves the best results for all evaluation metrics under both $\times2$ and $\times4$ downsampling rates. In Fig.~\ref{fig:chikusei-sr}, SHARE reconstructs images with significantly sharper textures, clearer structures and better spatial consistency. Compared to PnP-DIP and SSDL, SHARE not only delivers visually superior results but also shows improved spectral fidelity, as reflected in the lower SAM values. Please refer to the SM for a clearer comparison of high-resolution results.

\begin{figure}
    \centering
    \includegraphics[width=1\linewidth]{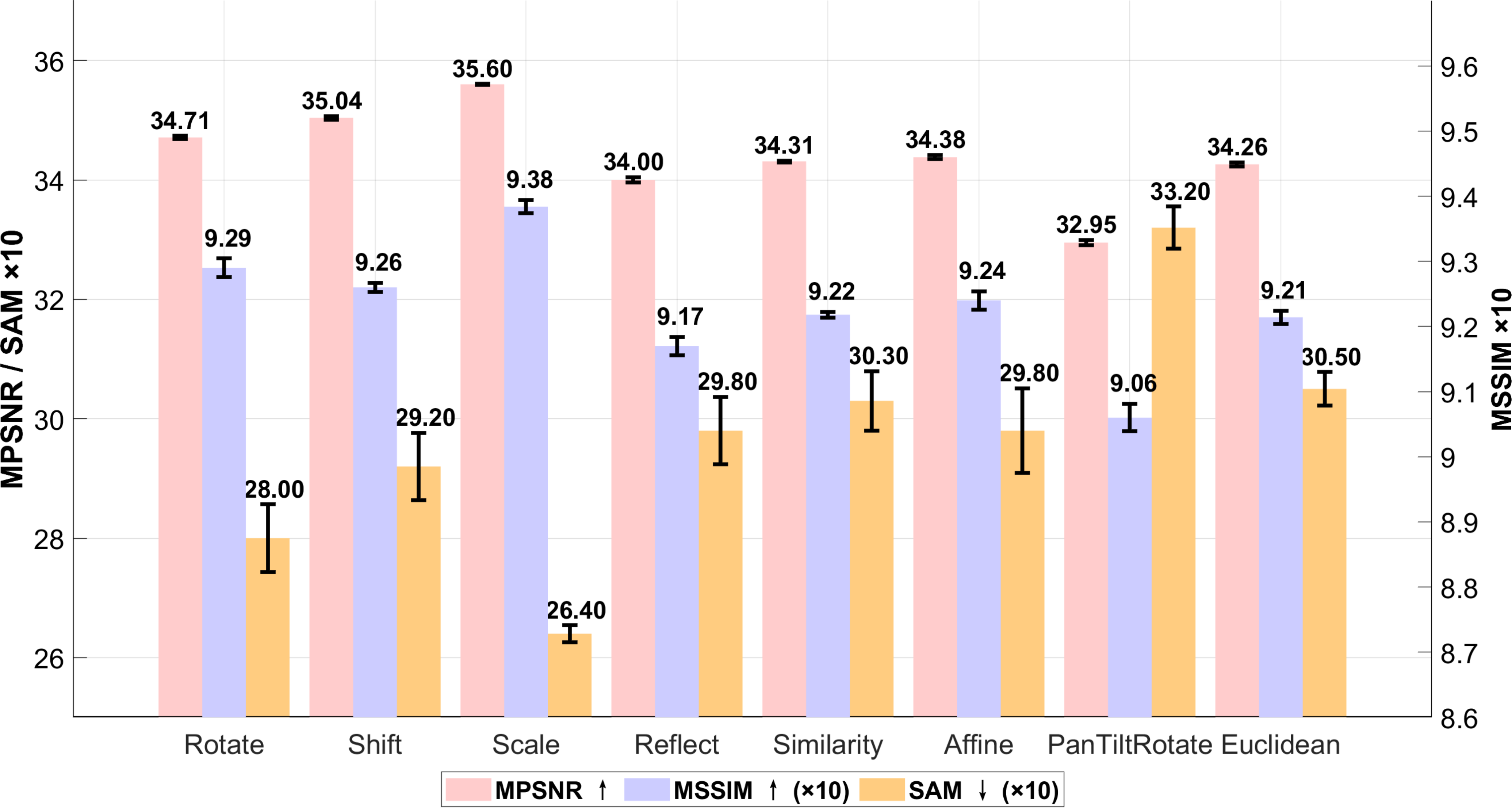}
    \caption{Comparison of different transformation results. \emph{Scale} achieves the best performance across all transformations and all metrics. We run each transformation experiment three times to draw the error bar.}
    \label{fig:transformation-result}
\end{figure}

\subsection{Ablation Study}

To verify the effectiveness of each core component in SHARE, we perform ablation studies on transformation selection, the DASA module, the $\alpha$ trade-off parameter, loss function terms, and different noise types and levels. The experiments are performed on the Cave fake and real Beers ms dataset with a $\times2$ downsampling rate.

\subsubsection{Transformation}
The selection of an appropriate transformation group is crucial for equivariant learning without ground truth \cite{SPM}. In this section we explore the effects of eight well-defined transformations, including four basic transformations: \emph{Rotation}, \emph{Shift}, \emph{Scaling}, \emph{Reflection}, and four projective transformations \cite{ei-pansharpen}: \emph{Similarity}, \emph{Affine}, \emph{PanTiltRotate} and \emph{Euclidean}. Please see SM for more details about each action. The quantitative results are shown in Fig.~\ref{fig:transformation-result}. Among them, the scale transformation achieves the best performance in the super-resolution task. Notably, all transformations except PanTiltRotate produce comparable and closely aligned results, indicating the inherent equivariance of the HSI data under these transformations and highlighting the benefits of learning with equivariance.
It is worth noting that some previous studies have demonstrated the effectiveness of specific group actions for particular inverse problems — e.g., Scale for super-resolution \cite{ei-sr} — and our results further support this conclusion.

\begin{figure}
    \centering
    \includegraphics[width=1\linewidth]{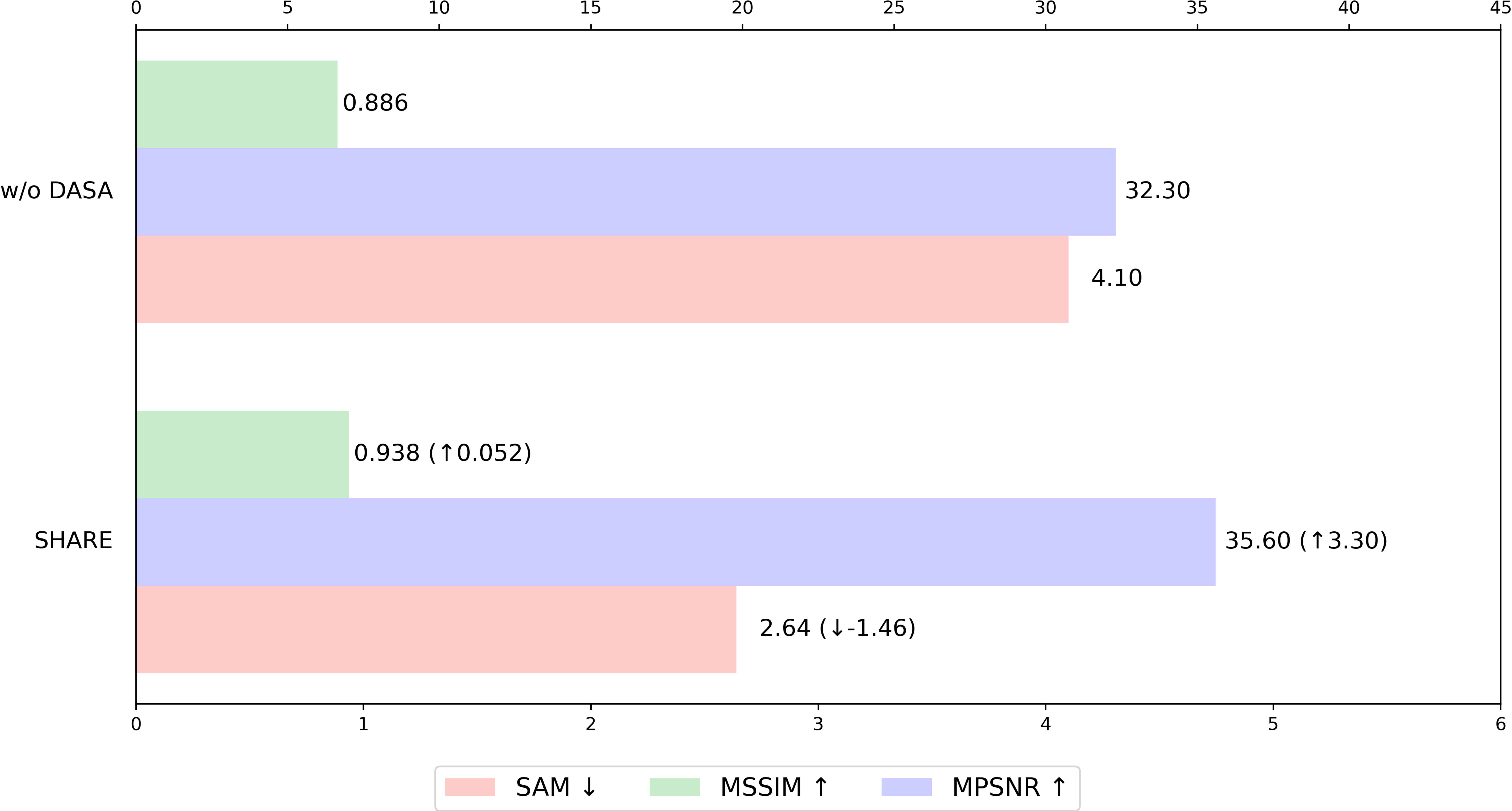}
    \caption{Ablation results of DASA. Our SHARE achieves the best performance when combined with DASA.}
    \label{fig:ablation-dasa}
\end{figure}

\begin{table}
    \centering
    \caption{Ablation analysis of the trade-off parameter $\alpha$. Large or small values of $\alpha$ lead to a decrease in SHARE performance. The best and second-best values are marked in \highlight{red}{red} and \highlight{blue}{blue}, respectively.}
    \small
    \begin{tabular}{cccccc}
    \toprule
    \textit{$\alpha$}  & $0.1$ & $0.5$ & $1$ & $1.5$ & $2$\\

    \midrule
    \textbf{MPSNR}$\uparrow$ & 31.93 & \highlight{blue}{34.47} & \highlight{red}{35.60} & 34.32
      & 34.02 \\
    \textbf{MSSIM}$\uparrow$ & 0.869 & 0.921 &  \highlight{red}{0.938} & \highlight{blue}{0.924}  & 0.916 \\
    \textbf{SAM}$\downarrow$   & 4.21 & 3.09 & \highlight{red}{2.64}  & \highlight{blue}{2.98}  & 3.05 \\
    \bottomrule
\end{tabular}

\label{tab:alpha}
\end{table}

\subsubsection{DASA}

In this section we remove the DASA module from $f_\theta$ and compare its performance with the full SHARE model. As shown in Fig.~\ref{fig:ablation-dasa}, the removal of DASA results in a decrease of 3.30 dB in MPSNR, 0.052 in MSSIM and an increase of 1.46 degrees in SAM.

\subsubsection{$\alpha$ Trade-off Parameter}
The $\alpha$ parameter weights $\mathcal{L}_{EC}$ and $\mathcal{L}_{sure}$. A large $\alpha$ will emphasize the learning of null space information, but underweight the importance of the range space signal. Conversely, a small $\alpha$ will limit range space reconstruction and make it difficult to recover the null space element. Therefore, $\alpha$ is a crucial hyperparameter in the SHARE training phase. The quantitative results are shown in Table~\ref{tab:alpha}. When $\alpha=1$, a balance is reached between $\mathcal{L}_{sure}$ and $\mathcal{L}_{ec}$.

\subsubsection{Loss Term}
Our SHARE evolves from standard MC (Eq.~\ref{eq:mc}) and vanilla equivariance constraint (Eq.~\ref{eq:ei}) to SURE (Eq.~\ref{eq:final-gaussian-sure}) and REC (Eq.~\ref{eq:share_rec}). To evaluate the effectiveness of the improved loss term, we run seven sets of experiments with different loss functions: 1) MC only; 2) SURE only; 3) REC only; 4) MC and EC; 5) SURE and EC; 6) MC and REC; 7) SURE and REC. Theoretically, MC or SURE alone primarily facilitates learning of range space components, whereas EC or REC alone tends to only recover null space information. Combinations other than SURE+REC may suffer from either noise sensitivity or unreliable null space reconstruction. The results in Table~\ref{tab:loss-term} clearly show the robust learning advantage of SURE and REC.

\begin{table*}
    \centering
    \caption{Ablation results of different loss function terms. Our SHARE achieves the best performance when $\mathcal{L}_{SURE}$ and $\mathcal{L}_{EC}$ are both incorporated. The best and second-best values are marked in \highlight{red}{red} and \highlight{blue}{blue}, respectively.}
    \small
    \begin{tabular}{lccccccc}
    \toprule
    \textit{terms}  & $t_1$ & $t_2$ & $t_3$ & $t_4$ & $t_5$ & $t_6$ & $t_7$\\
    \midrule
    $\mathcal{L}_{\text{mc}}$   & \ding{51} & w/o & w/o &\ding{51} & w/o &\ding{51} & w/o \\
    $\mathcal{L}_{\text{sure}}$  & w/o & \ding{51} & w/o & w/o & \ding{51} & w/o & \ding{51} \\
    $\mathcal{L}_{ec}$ & w/o & w/o & w/o & \ding{51} & \ding{51} & w/o & w/o\\
    $\mathcal{L}_{rec}$  & w/o & w/o & \ding{51} & w/o & w/o & \ding{51} & \ding{51}  \\

    \midrule
    \textbf{MPSNR}$\uparrow$ & 27.36  & 30.64 & 11.35 & 
     32.70 & 33.82 & \highlight{blue}{33.91}
      & \highlight{red}{35.60} \\
    \textbf{MSSIM} $\uparrow$ & 0.547 &  0.844& 0.052 & 0.898 & 0.914 & \highlight{blue}{0.918}  & \highlight{red}{0.938} \\
    \textbf{SAM} $\downarrow$  & 8.54 & 5.62 & 74.72  & 3.99 & 3.39 &\highlight{blue}{3.21}  & \highlight{red}{2.64} \\
    \bottomrule
\end{tabular}

\label{tab:loss-term}
\end{table*}

\subsubsection{Noise Types and Levels}

Due to space limitations, please see the SM for more details.

\begin{figure*}
    \centering
    \includegraphics[width=1\linewidth]{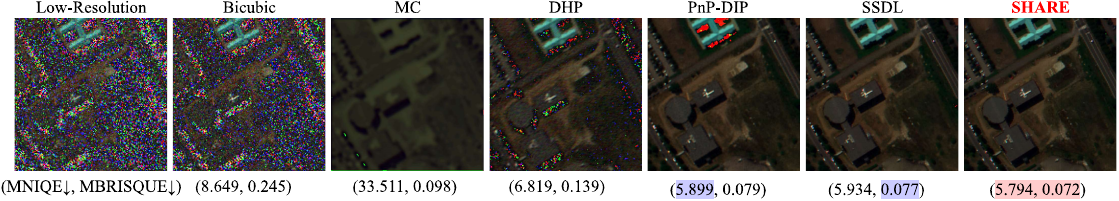}
    \caption{Comparison results of real-world $\times 2$ super-resolution. The low-resolution HSI is corrupted by Gaussian noise with a standard deviation of 25. MNIQE $\downarrow$ and MBRISQUE $\downarrow$ values are shown below each image. The best and second-best values are marked in \highlight{red}{red} and \highlight{blue}{blue}, respectively. Please zoom in for a better view.}
    \label{fig:real-sr}
\end{figure*}

\begin{table}
    \caption{Comparison of different methods training and testing on multiple \textbf{CAVE} HSI. The \highlight{red}{red} and \highlight{blue}{Blue} markers indicate the best and second-best values among all methods.}
    \centering
    \begin{tabular}{c|ccc}
        \toprule
        {Method}& MPSNR $\uparrow$ & MSSIM $\uparrow$ & SAM $\downarrow$   \\
        \cmidrule{1-4}
        Bicubic &  21.28 & 0.138 & 50.87  \\
        SSDL \cite{self-super-assisted-2024}& \highlight{blue}{31.57} & \highlight{blue}{0.842} & \highlight{blue}{14.39} \\
        MC &  29.42 & 0.632 & 26.05  \\
        \textbf{SHARE} & \highlight{red}{33.50} & \highlight{red}{0.880} & \highlight{red}{12.05} \\ 
        \bottomrule
    \end{tabular}
    \label{tab:sr-multi-cave}
\end{table}

\subsection{Multi-HSI Restoration}
In this section, we further demonstrate that SHARE can be used for multi-HSI image restoration tasks other than the restoration of a single HSI image. 
In particular, we conduct $\times 2$ super-resolution experiments on the CAVE dataset. The first 22 images are used as the training set and the remaining images form the test set. The other settings, such as the noise level and kernel size, align with those used in previous CAVE super-resolution experiments.

As shown in Table~\ref{tab:sr-multi-cave}, we have below observations: (i) Training SHARE with multiple measurements  (degraded images) yields markedly better reconstructions than training with a single measurement, and performance improves as more training images are provided. (ii) Higher resolution/quality training measurements (e.g. CAVE data) further boost accuracy. (iii) These gains hold in both the single-image and multi-image settings, demonstrating the strong scalability of SHARE across inpainting and super-resolution. This distinguishes it significantly from previous fully unsupervised methods \cite{DIP} \cite{DHIP}. Note high-quality datasets are not always available in practice. For example, satellite sequences may include frames with reduced spatial resolution or missing bands. The results show that SHARE remains applicable, as it can reliably recover corrupted information from even a single degraded image.

\subsection{Real-World Super-Resolution}
To comprehensively verify the effectiveness of our SHARE, we perform $\times 2$ real-world super-resolution experiments on the Pavia University dataset. We centrally crop a $160\times 160\times 103$ patch and then apply Gaussian noise with zero mean and standard deviation of 25 as a low-resolution HSI. To estimate the spatial response matrix $\Phi$ in Eq.~\ref{eq:sr-degrade}, we use the \cite{zsl} method by inputting a low-resolution HSI and a high-resolution MSI. We acquire the MSI by spectrally subsampling the reference image with the IKONOS-like spectral response.

We compare SHARE with bicubic interpolation, MC, DHP, PnP-DIP, SSDL, and evaluate the super-resolution performance using two no-reference image quality metrics: Mean NIQE (MNIQE) \cite{NIQE} and Mean BRISQUE (MBRISQUE) \cite{BRISQUE}. The results are shown in Fig.~\ref{fig:real-sr}. Our method, SHARE, consistently outperforms the other approaches on both metrics.

From a visual perspective, methods such as bicubic interpolation, DHP and PnP-DIP suffer from severe noise corruption. Although MC effectively removes most of the noise, the resulting images tend to be overly smooth and blurred, indicating a loss of high frequency detail. SSDL shows improvements in preserving features such as edges and patterns, but still introduces noticeable noise in the upper regions of the image. In contrast, SHARE successfully upsamples the hyperspectral images while preserving spatial fidelity and delivering more realistic and visually pleasing results.

\section{Conclusion}
\label{sec:conclusion}

We present SHARE, a novel fully unsupervised framework for single hyperspectral image (HSI) restoration that requires no ground truth data by effectively exploiting the inherent invariance properties of HSI. To address the critical challenge of noise in real-world sensing, SHARE integrates Stein's unbiased risk estimate (SURE) loss with robust equivariance consistency learning to achieve superior noise resilience. Our proposed dynamic adaptive spectral attention (DASA) module further enhances restoration accuracy at fine-grained levels. Extensive experiments demonstrate the state-of-the-art performance of SHARE in both HSI inpainting and super-resolution tasks. We hope that this work will open new directions for unsupervised HSI restoration and scientific imaging, with promising extensions to other challenging scenarios such as dehazing and deblurring.

\bibliography{reference} 
\bibliographystyle{IEEEtran}

\setcounter{page}{1}

\clearpage

\setcounter{page}{1}

\section*{Supplementary Material}
\subsection{More Inpainting Results}

\begin{figure*}
    \centering
    \includegraphics[width=1\linewidth]{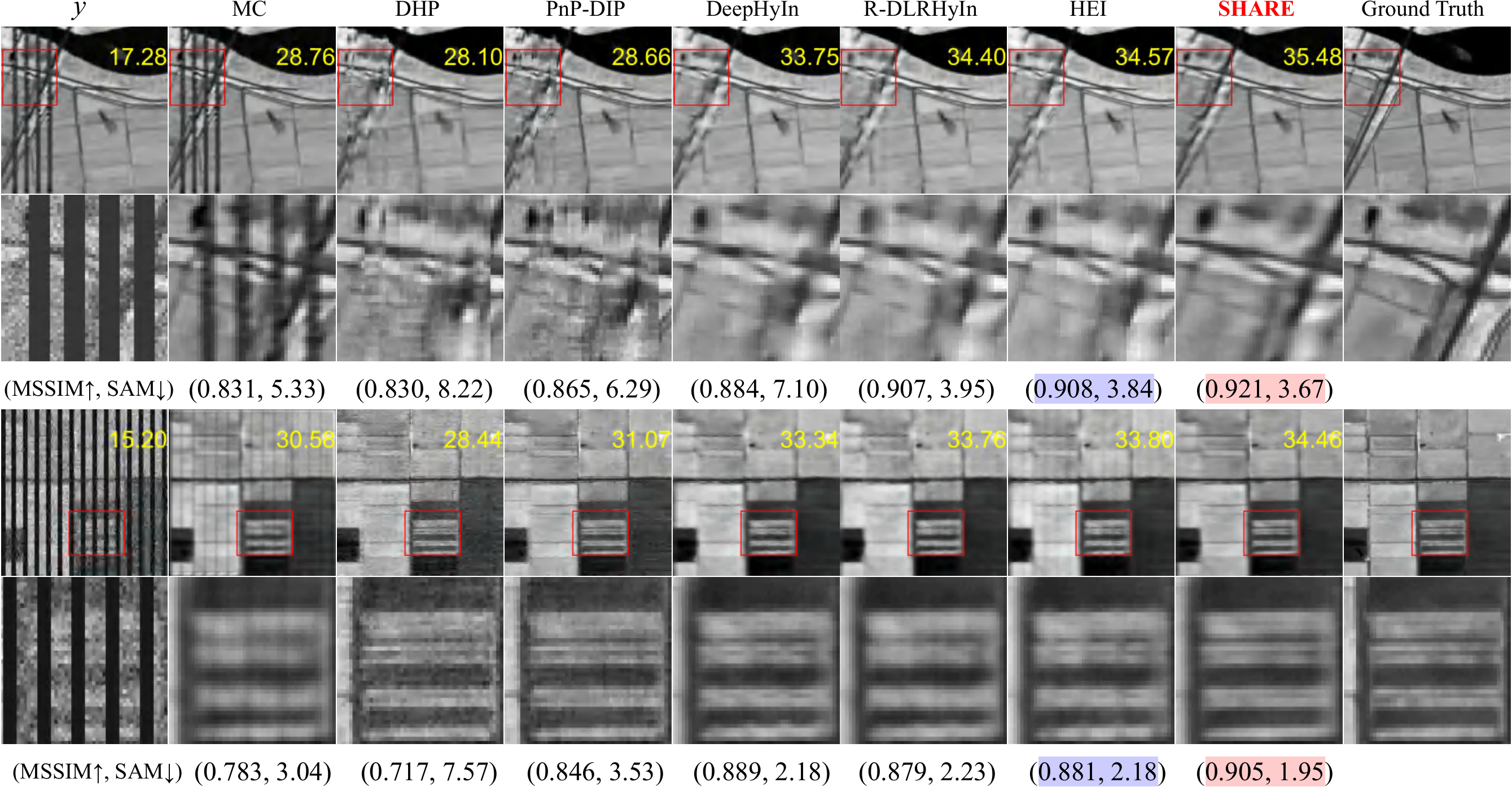}
    \caption{More inpainting results on CHikusei dataset with different mask shape.}
    \label{fig:inpainting-chikusei-sm}
\end{figure*}

\begin{figure*}
    \centering
    \includegraphics[width=1\linewidth]{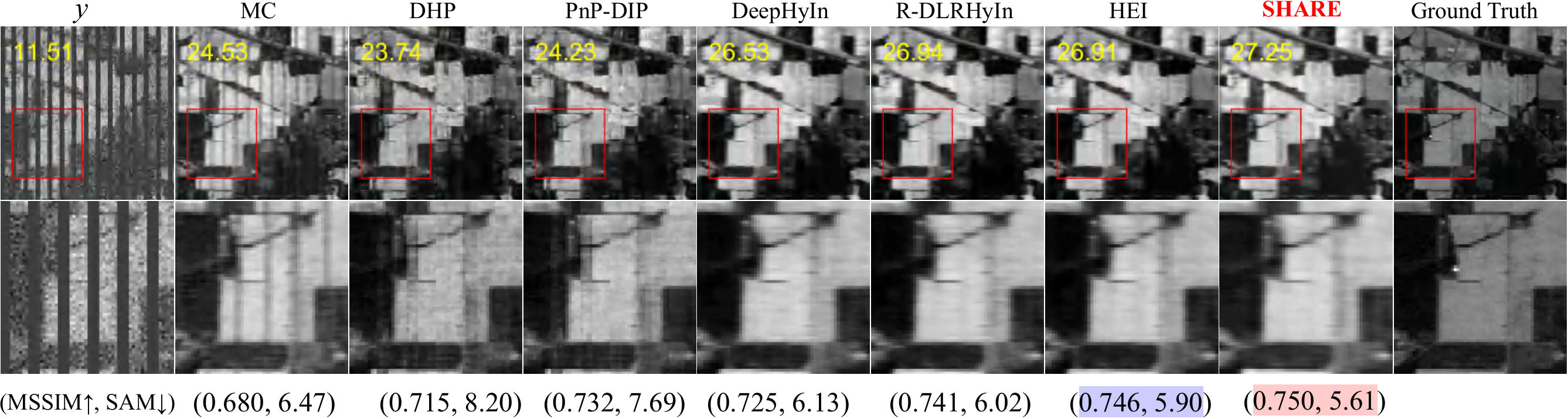}
    \caption{More inpainting results on Indian dataset with different mask shape.}
    \label{fig:inpainting-indian_sm}
\end{figure*}

Here we demonstrate more inpainting results on both Chikusei and Indian pines datasets with different mask shapes. The visualizations results are shown in Fig.~\ref{fig:inpainting-chikusei-sm} and Fig.~\ref{fig:inpainting-indian_sm}.
 
\subsection{Ablation Study}
\subsubsection{Transformation }

\begin{figure*}
    \centering
    \includegraphics[width=1\linewidth]{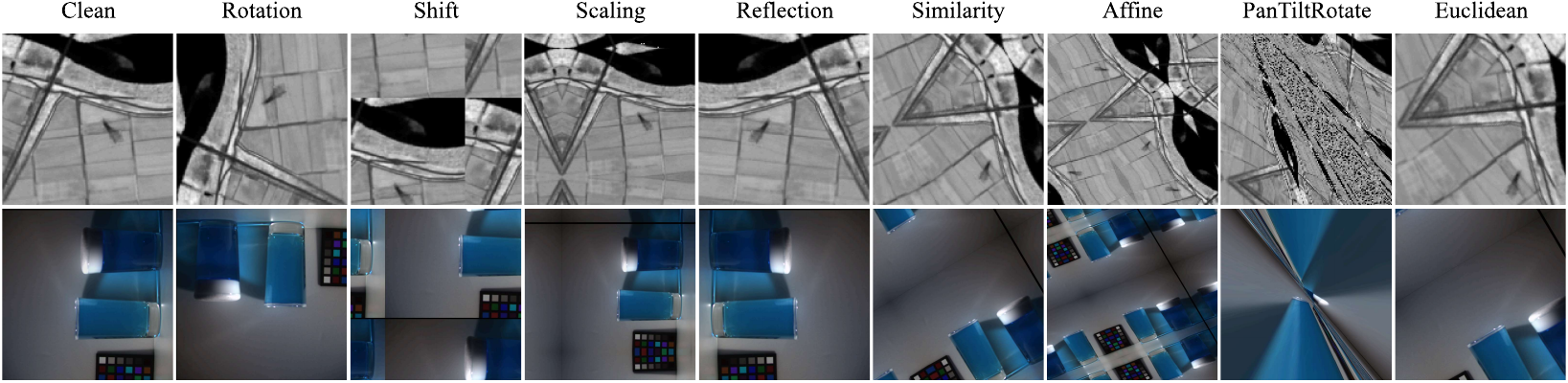}
    \caption{The visualization of each transformation result.}
    \label{fig:transformation-visual}
\end{figure*}

In the main text of the ablation study, we mentioned several geometry group actions. Here, we provide an explanation for each action:  Rotation, shift, scaling, reflection, and four projective transformations: Similarity, Affine, PanTiltRotate, and Euclidean. For clarity, we provide definitions of the projective transformations below: Similarity includes scaling, rotation and translation while preserving the aspect ratio and angles; this maintains shape similarity. Affine transformation allows translation, rotation, scaling, and shearing while keeping lines straight and parallel; however, it can potentially change proportions and angles. PanTiltRotate combines panning, tilting, and rotation to add perspective effects and create realistic viewpoints. The Euclidean transformation includes translation and rotation, preserving distances and angles without scaling. This makes it suitable for preserving rigid shapes. The visualisation results of each transformation are shown in Fig.~\ref{fig:transformation-visual}.

\subsubsection{Noise Types and Levels}



In real-world sensing conditions, observations are often distorted by various types and levels of noise. To comprehensively evaluate SHARE's robustness to noise, we performed six sets of experiments.

First, we added Gaussian noise with a zero mean and standard deviations of 25 and 50, respectively. Next, we tested SHARE under Poisson noise with gains of $1/25$ and $1/10$. Finally, we apply mixed Gaussian–Poisson noise using the same settings as in the previous groups.

\begin{figure*}
    \centering
    \includegraphics[width=1\linewidth]{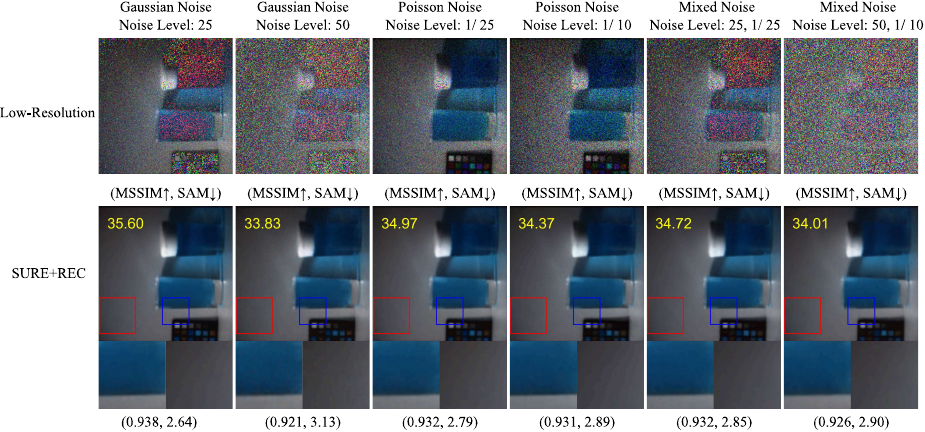}
    \caption{Comparison of different loss function combination results under different types and levels of noise. MPSNR values are shown in the top left corner, and MSSIM, SAM values are shown below each image. Please zoom in for a better view.}
    \label{fig:noise}
\end{figure*}

The visual and quantitative results are shown in Fig.~\ref{fig:noise}. SURE+REC achieves consistently better results, with only a slight drop in performance in the presence of severe noise for all settings.

\subsection{High-Resolution Experimental Results}

\begin{figure*}
    \centering
    \includegraphics[width=1\linewidth]{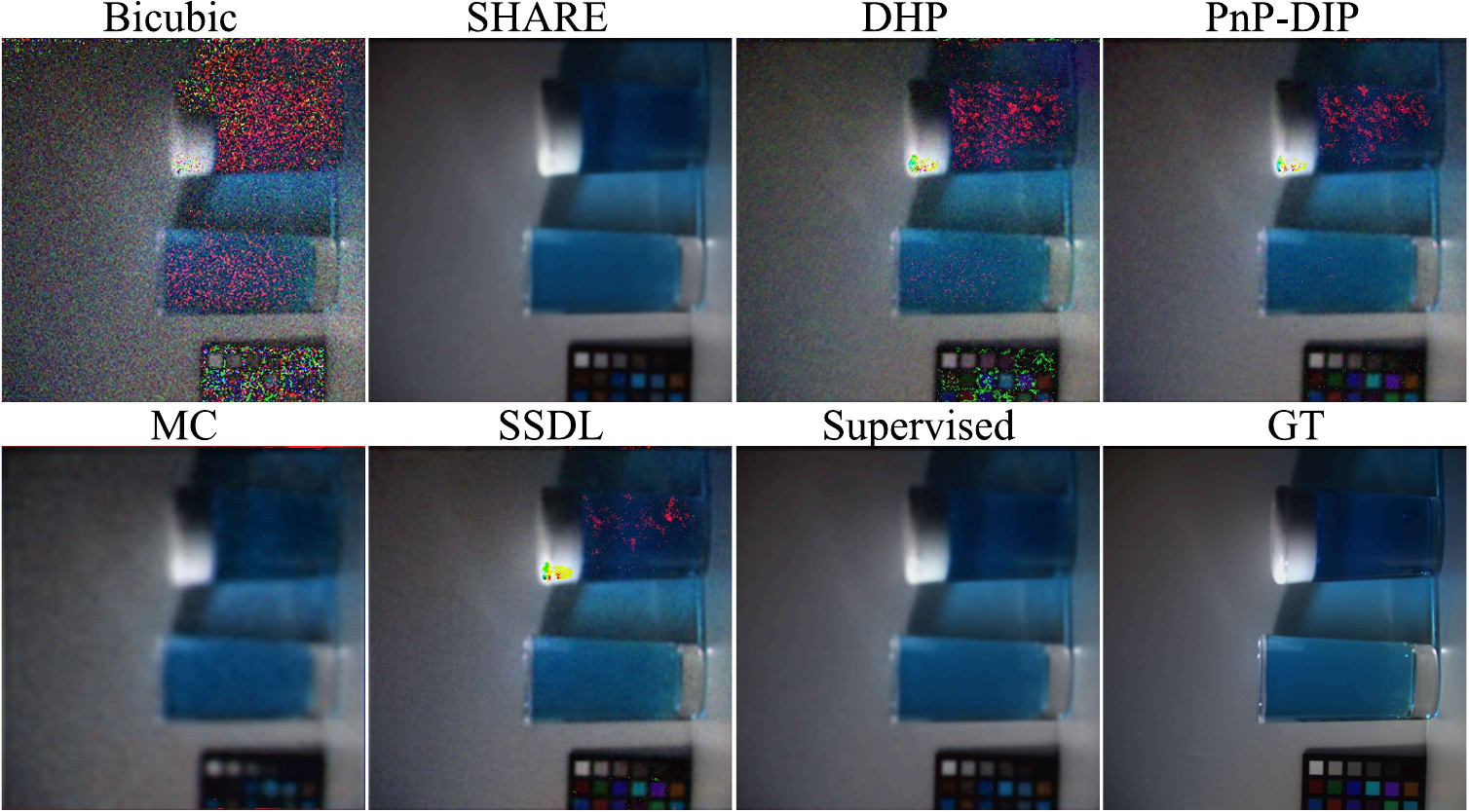}
    \caption{The high-resolution super-resolution results on \textbf{Cave: fake and real beers ms} dataset under $\times 2$ dowmsampling.}
    \label{fig:beers-HR-x2}
\end{figure*}

\begin{figure*}
    \centering
    \includegraphics[width=1\linewidth]{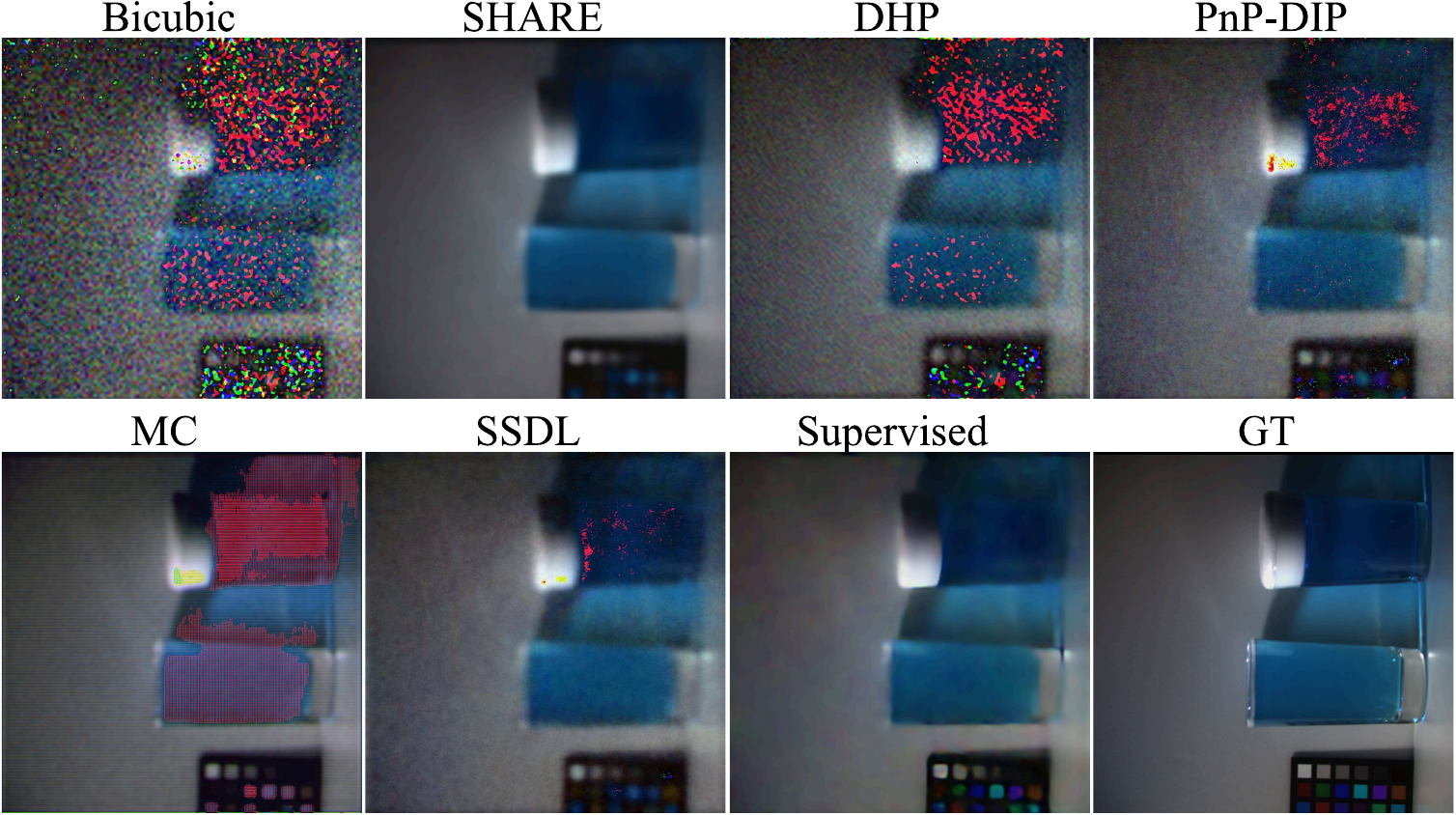}
    \caption{The high-resolution super-resolution results on \textbf{Cave: fake and real beers ms} dataset under $\times 4$ dowmsampling.}
    \label{fig:beers-HR-x4}
\end{figure*}

\begin{figure*}
    \centering
    \includegraphics[width=1\linewidth]{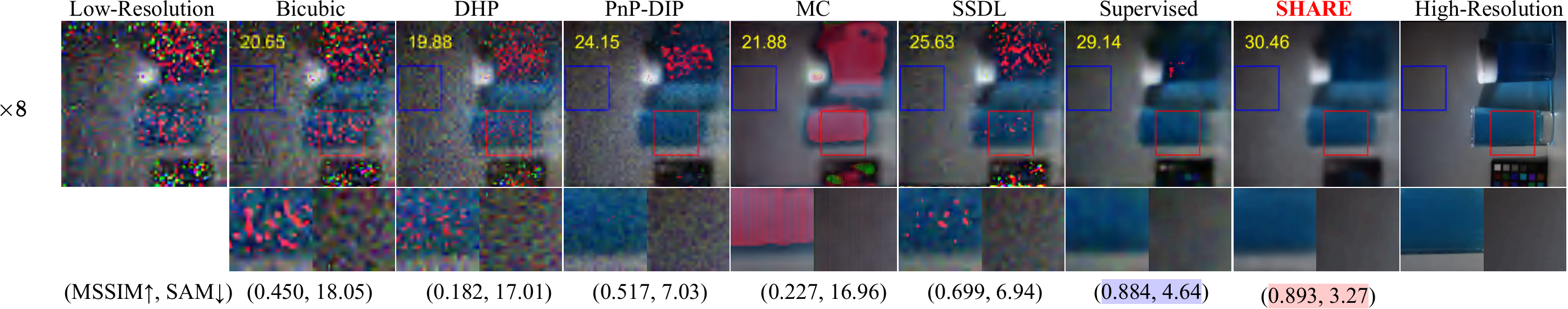}
    \caption{Super resolution results of different methods on \textbf{Cave: fake and real beers ms} dataset. MPSNR values are shown on the top left, MSSIM and SAM values are shown below each image. The best and second-best values are marked in \highlight{red}{red} and \highlight{blue}{blue}, respectively. Bands 5, 15, 25 for visualization. Please zoom in for a better view.}
    \label{fig:beers-SM}
\end{figure*}

\begin{figure*}
    \centering
    \includegraphics[width=1\linewidth]{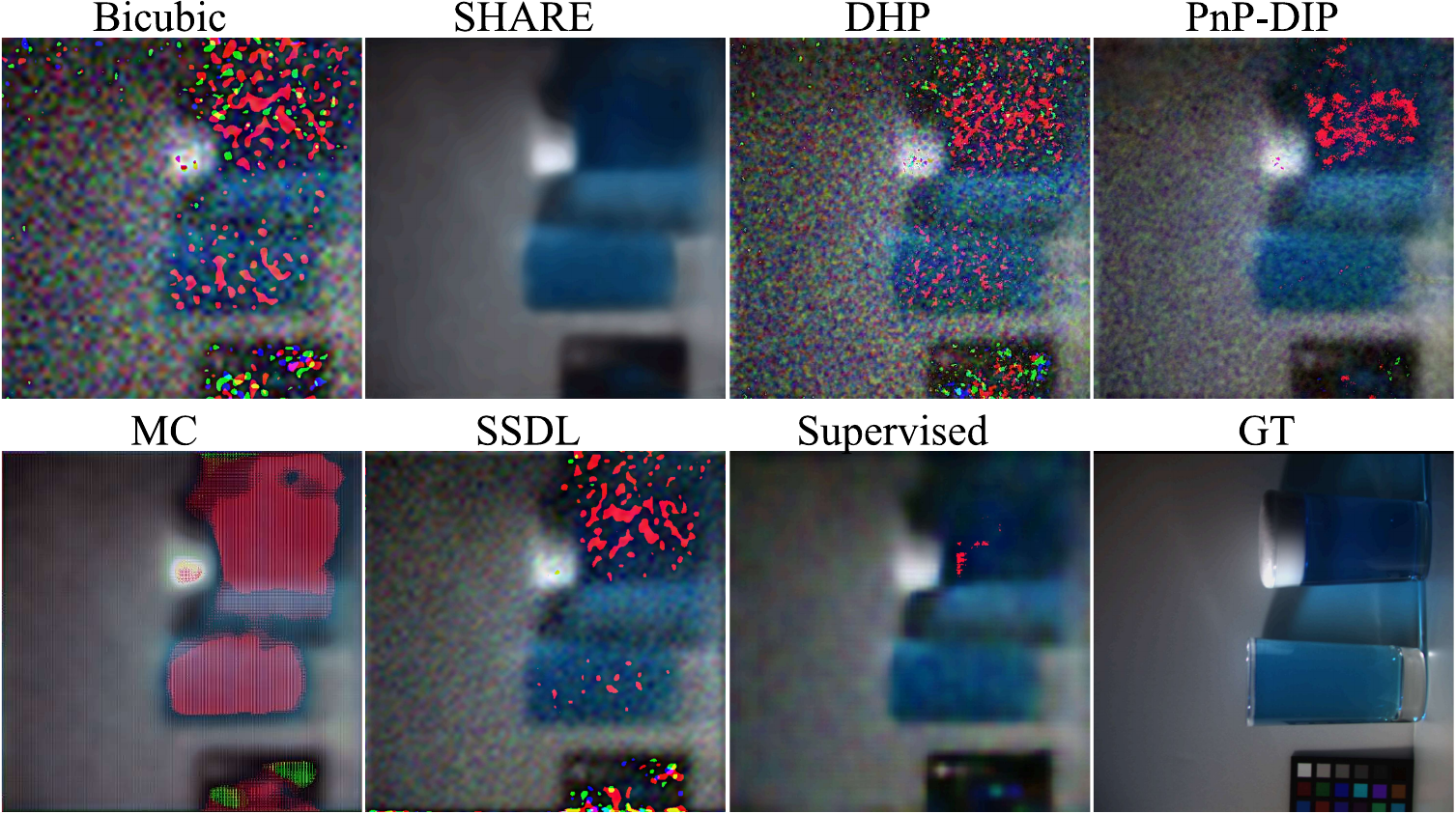}
    \caption{The high-resolution super-resolution results on \textbf{Cave: fake and real beers ms} dataset under $\times 8$ dowmsampling.}
    \label{fig:beers-HR-x8}
\end{figure*}

\begin{figure*}
    \centering
    \includegraphics[width=1\linewidth]{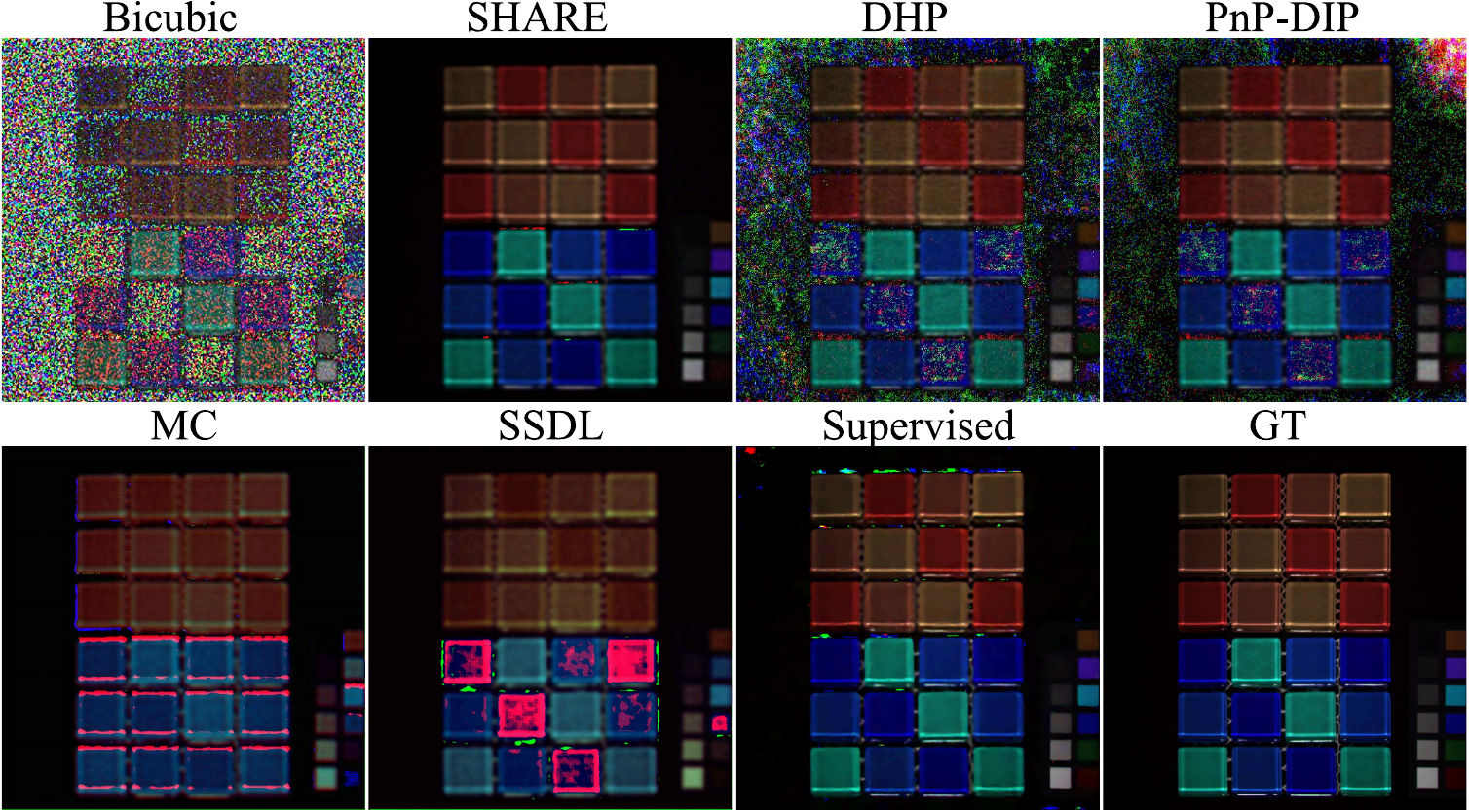}
    \caption{The high-resolution super-resolution results on \textbf{Cave: glass tiles ms} dataset under $\times 2$ dowmsampling.}
    \label{fig:glass-HR-x2}
\end{figure*}

\begin{figure*}
    \centering
    \includegraphics[width=1\linewidth]{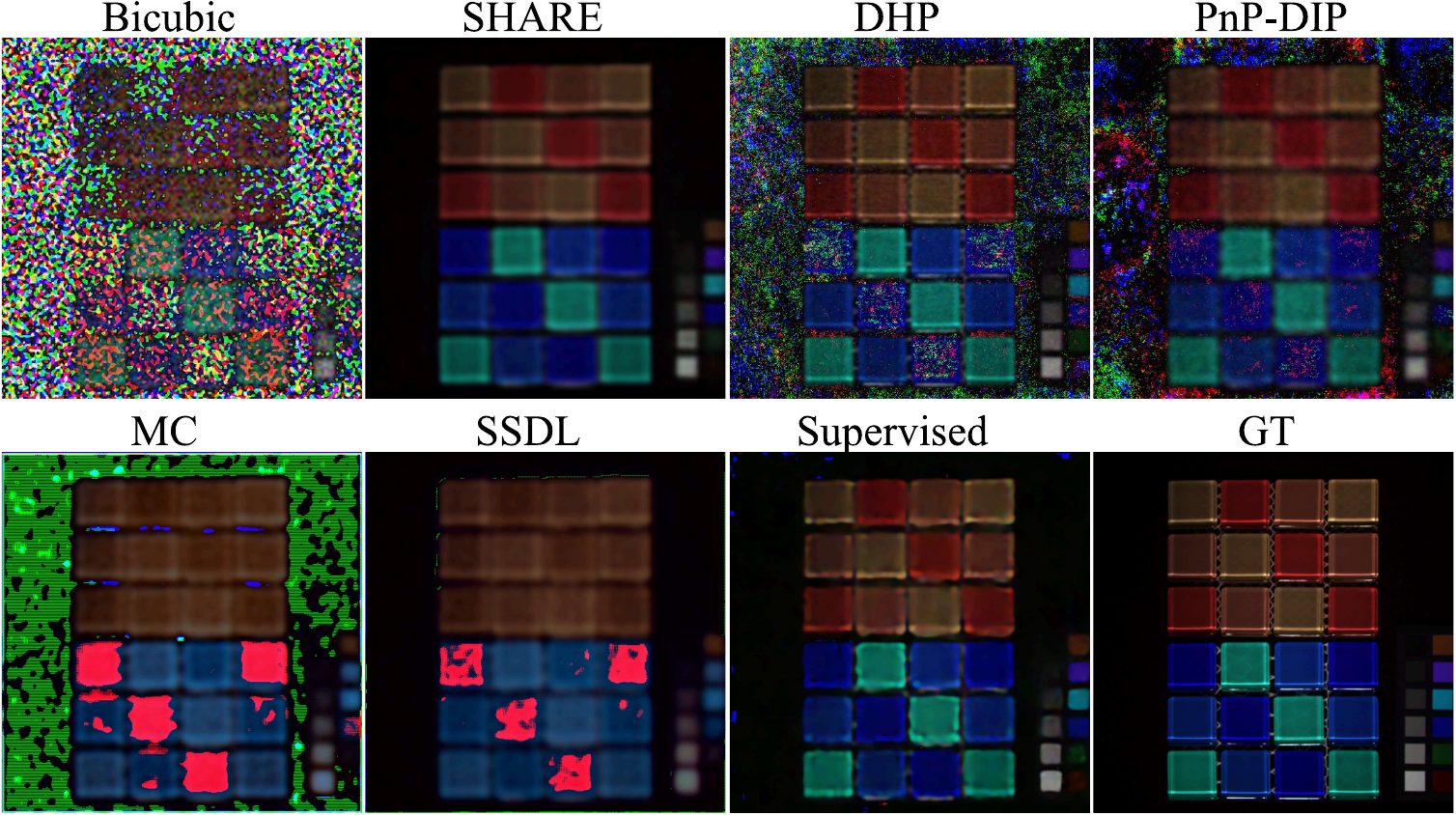}
    \caption{The high-resolution super-resolution results on \textbf{Cave: glass tiles ms} dataset under $\times 4$ dowmsampling.}
    \label{fig:glass-HR-x4}
\end{figure*}

\begin{figure*}
    \centering
    \includegraphics[width=1\linewidth]{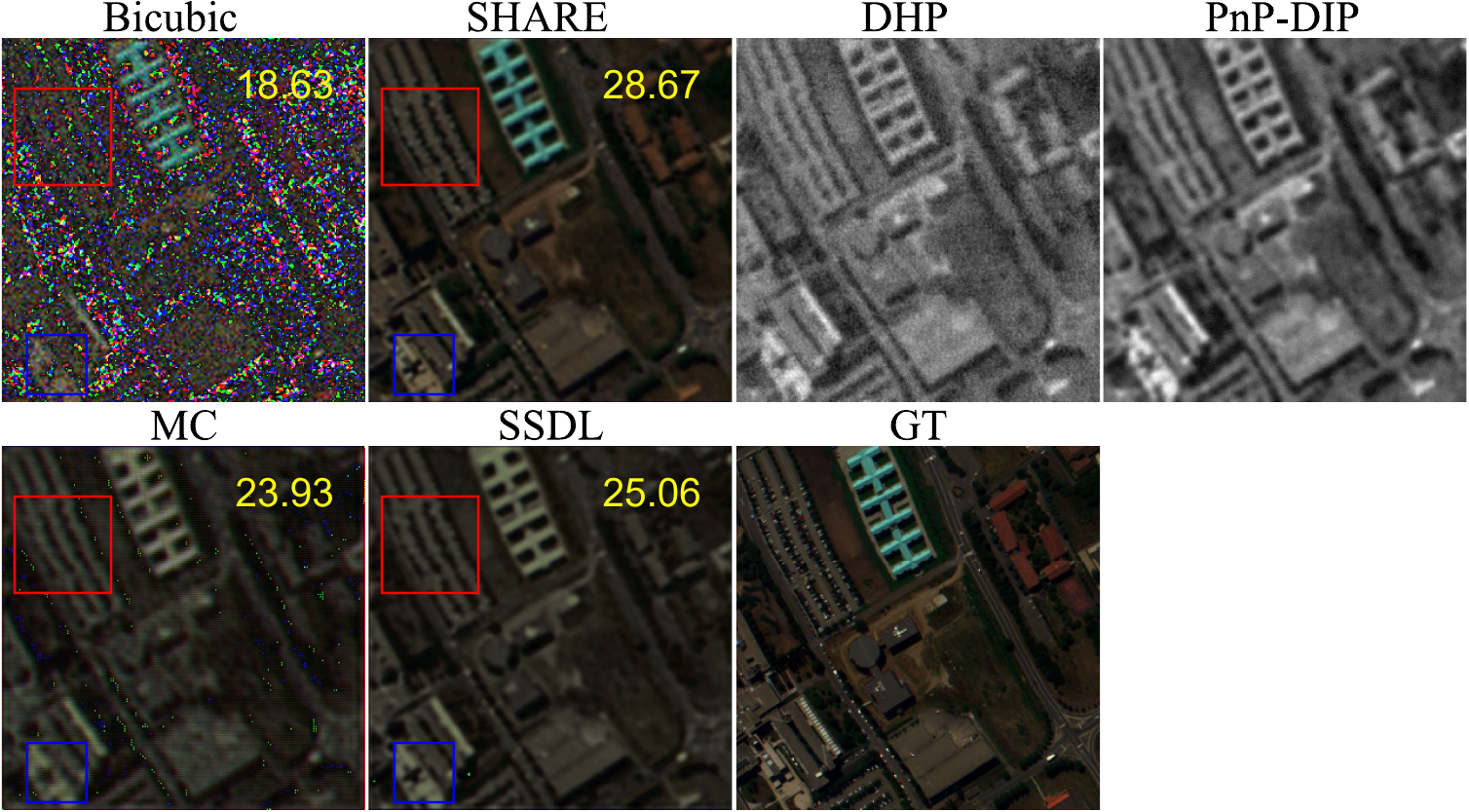}
    \caption{The high-resolution super-resolution results on \textbf{Pavia University} dataset under $\times 2$ dowmsampling.}
    \label{fig:pavia-HR-x2}
\end{figure*}

\begin{figure*}
    \centering
    \includegraphics[width=1\linewidth]{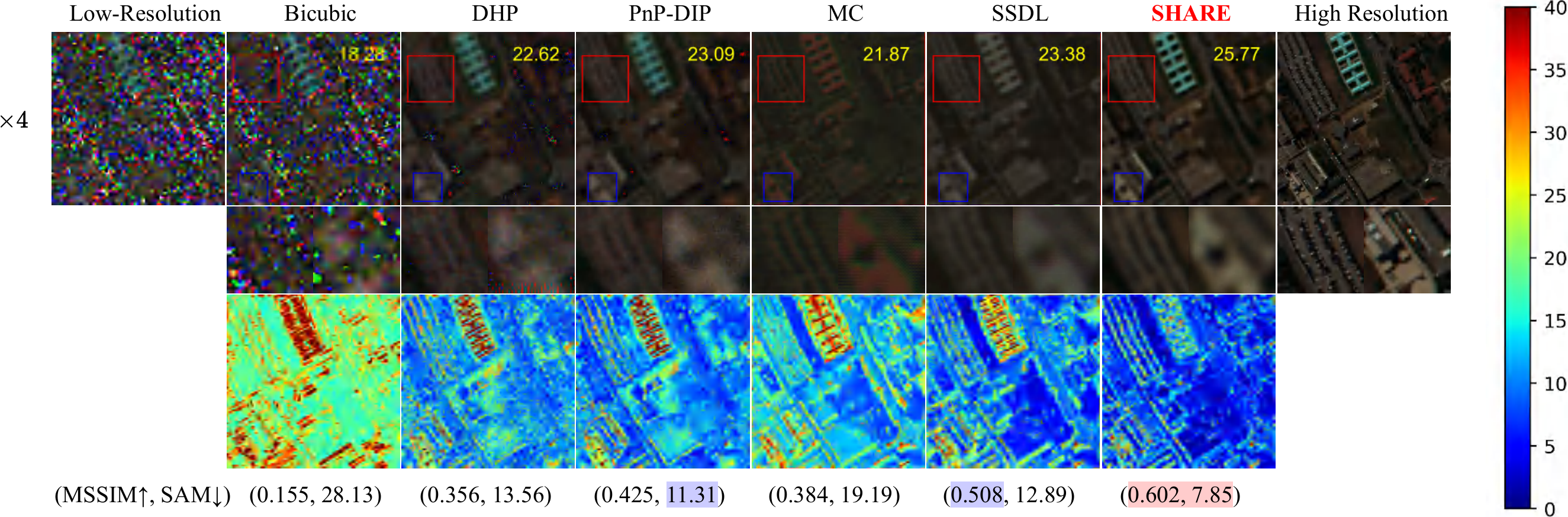}
    \caption{Super resolution results of different methods on \textbf{Pavia University} dataset. MPSNR values are shown on the top left, MSSIM and SAM values are shown below each image. The best and second-best values are marked in \highlight{red}{red} and \highlight{blue}{blue}, respectively. Bands 60, 29, 7 for visualization. First row: Super resolution results. Second row: Local details. Third row: Absolute error map. Please zoom in for a better view.}
    \label{fig:pavia-SM}
\end{figure*}

\begin{figure*}
    \centering
    \includegraphics[width=1\linewidth]{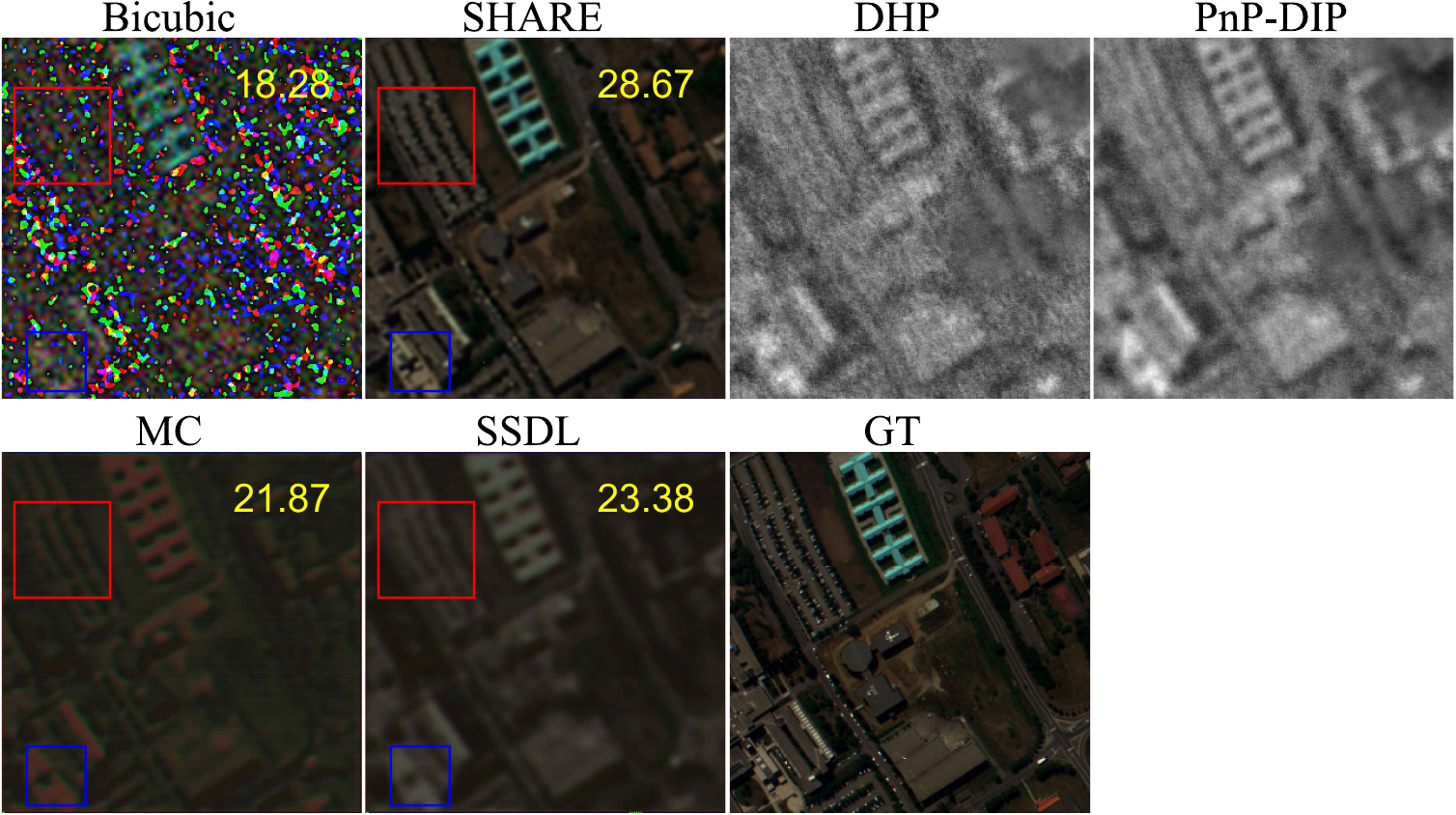}
    \caption{The high-resolution super-resolution results on \textbf{Pavia University} dataset under $\times 4$ dowmsampling.}
    \label{fig:pavia-HR-x4}
\end{figure*}

\begin{figure*}
    \centering
    \includegraphics[width=1\linewidth]{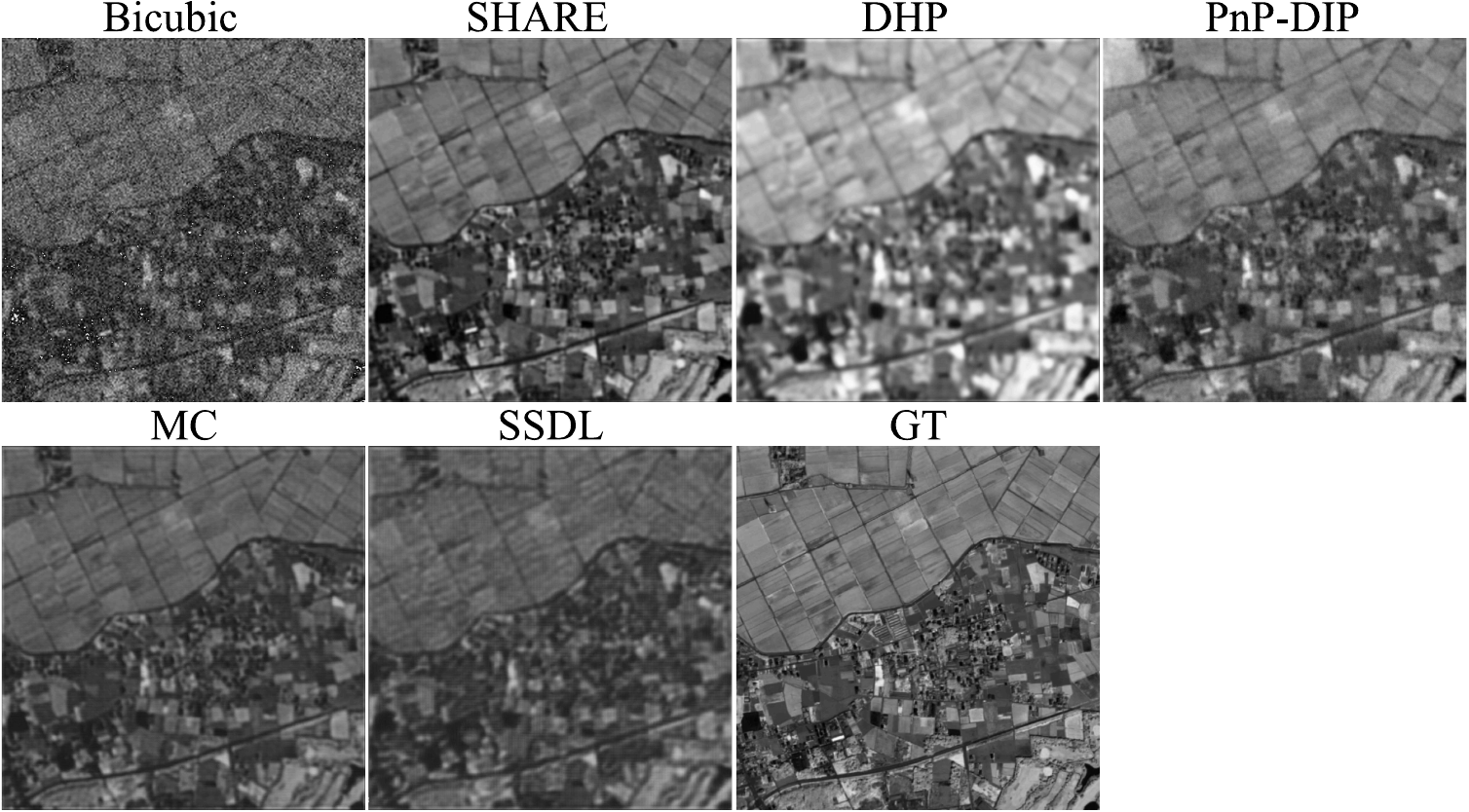}
    \caption{The high-resolution super-resolution results on \textbf{Chikusei} dataset1 under $\times 2$ dowmsampling.}
    \label{fig:chikusei-1-x2}
\end{figure*}

\begin{figure*}
    \centering
    \includegraphics[width=1\linewidth]{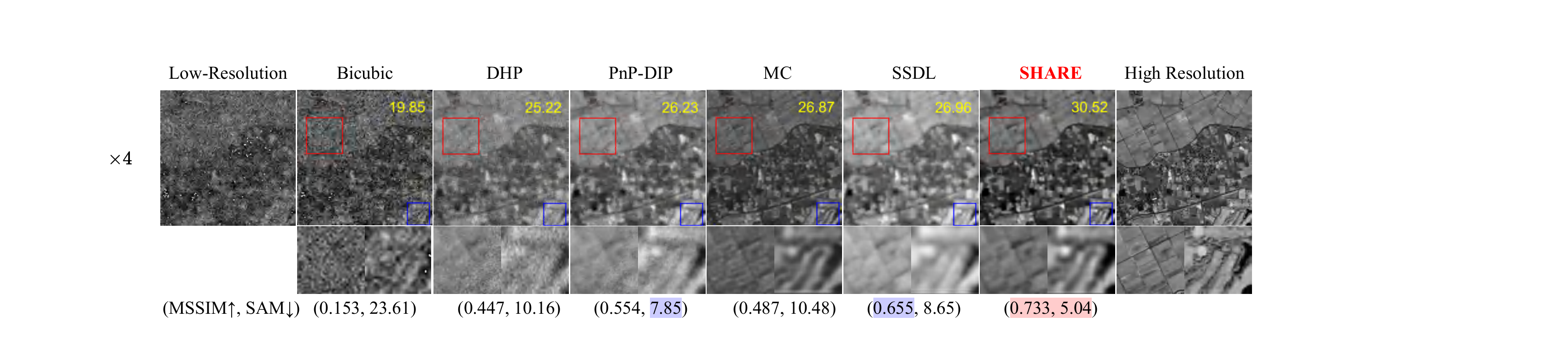}
    \caption{Super resolution results of different methods on \textbf{Chikusei} dataset. We crop two subfigures from Chikusei. MPSNR values are shown on the top left, MSSIM and SAM values are shown below each image. The best and second-best values are marked in \highlight{red}{red} and \highlight{blue}{blue}, respectively. Band 90 for visualization. Please zoom in for a better view.}
    \label{fig:chikusei-sr-SM}
\end{figure*}

\begin{figure*}
    \centering
    \includegraphics[width=1\linewidth]{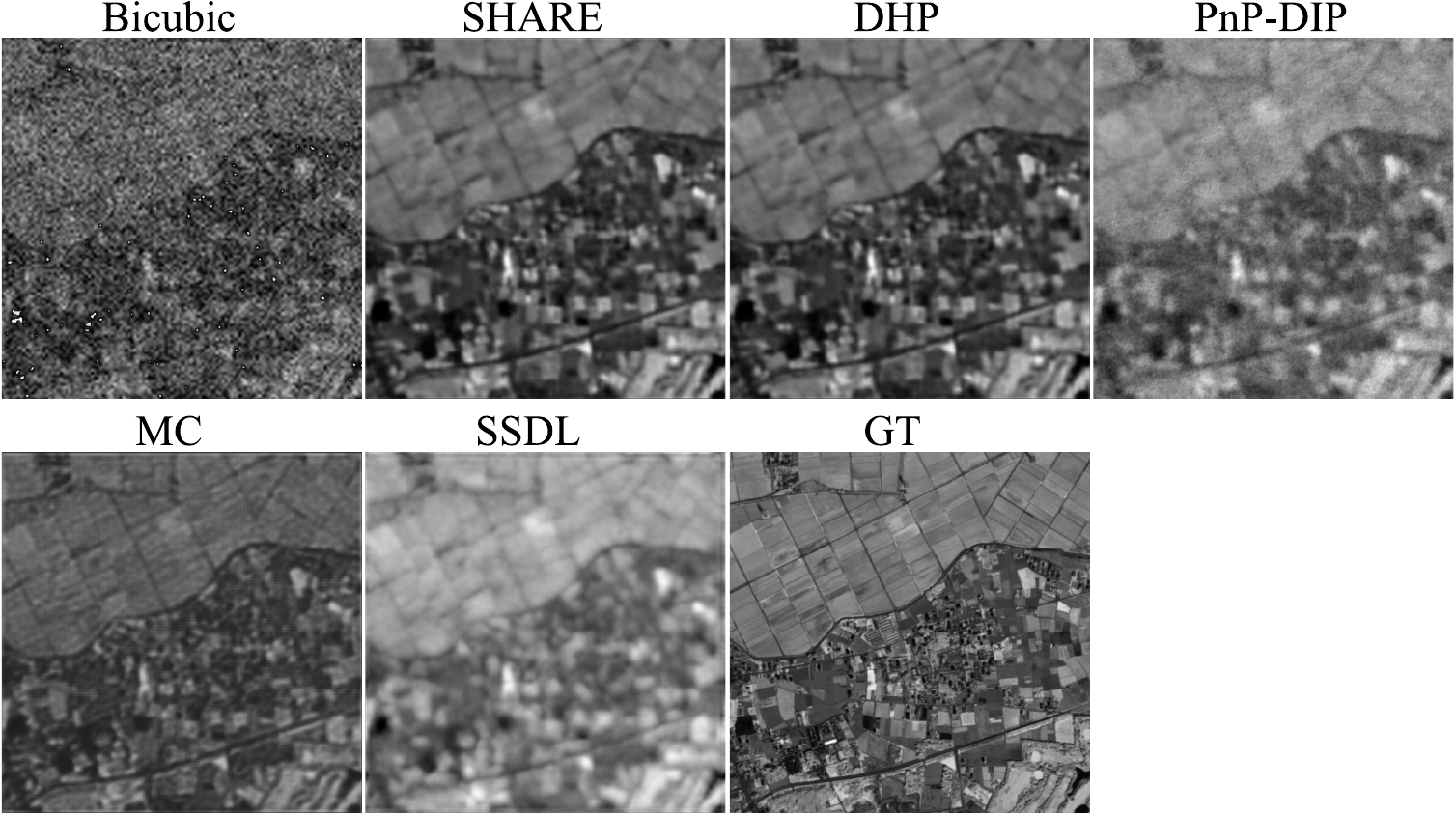}
    \caption{The high-resolution super-resolution results on \textbf{Chikusei} dataset1 under $\times 4$ dowmsampling.}
    \label{fig:chikusei-1-x4}
\end{figure*}

Due to space limitations in the main text, the super-resolution results were presented in a relatively low-resolution format. Here, we provide high-resolution figures (Fig.~\ref{fig:beers-HR-x2} - Fig.~\ref{fig:chikusei-1-x4}) to better illustrate the effectiveness of our SHARE method, allowing finer details to be more clearly observed.

\end{document}